\documentclass[12pt]{iopart}
\usepackage{subcaption}
\usepackage{xcolor}
\usepackage[
backend=biber,
style=phys,
sorting=none
]{biblatex}
\usepackage{graphicx,color}

\usepackage{iopams}  
\begin{document}

\title[]{Exploring symmetries in photoelectron holography with two-color linearly polarized fields}

\author{T. Rook and C. Figueira de Morisson Faria}

\address{Department of Physics and Astronomy, University College London\\ Gower Street, London WC1E 6BT, UK}
\ead{c.faria@ucl.ac.uk}
\vspace{10pt}
\date{\today}

\begin{abstract}
We investigate photoelectron holography in bichromatic linearly polarized fields of commensurate frequencies $r\omega$ and $s\omega$, with emphasis on the existing symmetries and for which values of the relative phase between the two driving waves they are kept or broken. Using group-theoretical methods, we show that, additionally to the well-known half-cycle symmetry, which is broken for $r+s$ odd, there are reflection symmetries around the field zero crossings and maxima, which may or may not be kept, depending on how both waves are dephased. The three symmetries are always present for monochromatic fields, while for bichromatic fields this is not guaranteed, even if $r+s$ is even and the half-cycle symmetry is retained. Breaking the half-cycle symmetry automatically breaks one of the other two, while, if the half-cycle symmetry is retained, the other two symmetries are either \textit{both} kept or broken. We analyze how these features affect the ionization times and saddle-point equations for different bichromatic fields. We also provide general expressions for the relative phases $\phi$  which retain specific symmetries. 
As an application, we compute photoelectron momentum distributions for $\omega-2\omega$ fields with the Coulomb Quantum Orbit Strong-Field approximation and assess how holographic structures such as the fan, the spider and interference carpets behave, focusing on the reflection symmetries. The features encountered can  be traced back to the field gradient and amplitude affecting ionization probabilities and quantum interference in different momentum regions.
\end{abstract}

\section{Introduction}
Resolving and steering electron dynamics in real time are key objectives of attosecond science \cite{Lein2007,Krausz2009,Salieres2012R,Gallmann2012}. In order to realize these aims, several research avenues have been pursued, such as attosecond pulses \cite{Agostini2004,Calegari2016}, high-order harmonic spectroscopy \cite{Marangos2016}, and ultrafast photoelectron holography \cite{Faria2020}. Due to the high intensities involved, the external laser field dictates the time scales and the dynamics. This is a consequence of strong-field phenomena being laser-induced processes, in which an electron is freed from its parent ion, propagates in the continuum and either reaches the detector or is brought back by the field to its parent ion, with which it may recombine or recollide \cite{Corkum1993}. Recollision leads to high-energy photoelectrons in above-threshold ionization (ATI) \cite{Becker2018,Becker2002Review,MilosReviewATI} or nonsequential double and multiple ionization (NSDI, NSMI) \cite{Faria2011,Becker2012}, while recombination results in high-order harmonic generation (HHG) \cite{Lewenstein1994}. 
For that reason, tailored fields have been widely explored as subfemtosecond-control tools for over three decades (for reviews see, e.g.,  \cite{Brabec2000,Ehlotzky2001,Milos2006}). This interest led to a multitude of applications, such as the \textit{in situ} characterization of attosecond pulses \cite{Dudovich2006,Doumy2009,Dahl2009}, the measurement of tunneling times \cite{Eckle2008,Pfeiffer2011,Pfeiffer2012,Shafir2012,Li2013,Zhao2013,Ivanov2014,Pedatzur2015,Han2017b,Landsman2013,Torlina2015,Henkel2015,Klaiber2018,Eicke2019}, probing chiral systems \cite{Smirnova2015JPhysB,Ayuso2018,Ayuso2018II,Baykusheva2018}, angular features in HHG, vortex-type interference \cite{Maxwell2021} and the phase-of-the phase spectroscopy using collinear \cite{Skruszewicz2015,Almajid2017,Wuerzler2020} or circularly polarized \cite{Tulsky2018,Tulsky2020} two-color fields. Apart from the usual linearly polarized bichromatic fields \cite{Ehlotzky2001} and few-cycle pulses \cite{Brabec2000}, elliptically polarized fields \cite{Landsman2014}, orthogonally polarized two-color (OTC)  \cite{Zhao2013,Zhang2014,Richter2015,Das2013,Das2015,Henkel2015,Li2016,Han2017,Gong2017,Xie2017,Habibovich2021} and bicircular  \cite{Milos2000,Smirnova2015JPhysB,Milos2016,Mancuso2016,Hoang2017,Eckart2018,Milos2018,Ayuso2018,Ayuso2018II,Baykusheva2018,Eicke2019,Yue2020,Maxwell2021} fields have been proposed and studied, as well as chiral \cite{Rozen2019} and knotted fields \cite{Pisanty2019}.

Thereby, dynamical symmetries \cite{Busuladvic2017,Neufeld2019,Yue2020,Habibovich2021} play a huge role, and may stem from the field polarization, time profile, relative frequencies and relative phases. Symmetries have been used, together with properties of the target, to determine selection rules in a wide range of scenarios. Besides the seminal work in \cite{Alon1998}, in recent years selection rules for HHG \cite{Milos2015} and strong-field ionization \cite{Busuladvic2017} in bicircular fields have been derived.  Further studies have focused on the role of the orbital angular momentum (OAM) in photoelectron vortices  \cite{NgokoDjiokap2015,Bayer2020,Kang2021,Maxwell2021}, molecules \cite{Liu2016,Yue2020}, and strong-field ionization in circularly polarized fields \cite{Barth2011,Barth2013}. This builds up on early work, which shows that HHG with two-color fields are dependent on the target \cite{Frolov2010}, and that an electron's angle of return will manifest itself as dynamic shifts in structural minima in HHG from diatomic targets \cite{Kitzler2008,Das2013,Das2015}. 

A two-colour field of linear polarization, composed of a wave of frequency $\omega$ and its second harmonic, delayed by a relative phase $\phi$, is among the simplest tailored fields. Since the 1990s, they have been widely used to break the half-cycle symmetry of a monochromatic wave. This symmetry means that, for a time translation $t \rightarrow t+T/2$, where $T$ is the field cycle, the electric field and corresponding vector potential will not change apart from a spatial reflection in the plane perpendicular to the polarization axis, that is, $E(t) =-E(t+T/2)$ and $A(t)=- A(t+T/2)$. Hence, for monochromatic driving fields or long enough pulses, photoelectron momentum distributions will be symmetric with regard to momentum reflections $p_{\parallel} \rightarrow -p_{\parallel}$ and $p_{\perp}\rightarrow -p_{\perp}$, where $p_{\parallel}$  and $p_{\perp}$ are the momentum components parallel and perpendicular to the field polarization axis. Inclusion of a second harmonic wave will break the symmetry with regard to $p_{\parallel}$. For high-harmonic spectra, breaking the half-cycle symmetry with an $\omega - 2\omega$ field will lead to even harmonics, which can be manipulated by altering the relative phase $\phi$. Breaking this symmetry, together with the phase dependence, has been hugely important for a wide range of applications, such as attosecond-pulse characterization \cite{Dudovich2006}, determining ionization times \cite{Zhao2013,Porat2018,Eicke2019} and the phase-of the phase spectroscopy \cite{Skruszewicz2015,Almajid2017,Tulsky2018,Wuerzler2020,Tulsky2020}. Often, a weak $2\omega$ wave is employed to minimally disrupt the continuum dynamics determined by the low frequency wave. Stronger $2\omega$ fields will alter the electron propagation in the continuum, and lead to a double plateau  \cite{Faria1999,Faria2000,Faria2001} or caustic-type structures \cite{Raz2012,Strelkov2016,Facciala2016,Hamilton2017} in high-order harmonic generation. Another way of breaking this symmetry is to use few-cycle pulses (see \cite{Shvetsov2014,Oi2017,López2019A,López2019B} for photoelectron holography). In contrast, a linearly polarized $\omega-3\omega$ field does not break the half cycle symmetry. One should note, however, that a linearly polarized bichromatic field may exhibit other, subtler symmetries, which have been studied in lesser depth. Examples are temporal reflections with regard to the field maxima or zero crossings, which depend on relative phases and frequency ratio between the two driving waves. 

In the present work, we will investigate what symmetries exist for linearly polarized two-color fields, under what conditions they are broken and what effects they have on specific holographic patterns. Examples are the fan-shaped fringes that form near the ionization threshold 
\cite{Rudenko2004,Maharjan2006,Gopal2009,Chen2006,Arbó2006}, the spider-like fringes that form near the polarization axis \cite{HuismansScience2011,Bian2011,Huismans2012,Marchenko2011,Hickstein2012,Moeller2014,Meckel2014},
and the spiral-like structure that forms perpendicular to the field-polarization axis and, in the high-energy region, leads to interference carpets \cite{Maxwell2020,Qin2021}.
We will focus on how their contrast and prominence depends on such symmetries and how they can be manipulated by altering the field parameters. Theoretical and experimental studies show that these features change for orthogonally \cite{Xie2015,Gong2017,Han2017,Han2018} 
and linearly polarized 
\cite{Xie2012,Xie2013,Arbo2014,Skruszewicz2015,Arbó2015,Xie2016b,Luo2017,Porat2018} 
two-color fields. 

Furthermore, in order to model the holographic patterns and assess how they form, one must employ an orbit-based method that incorporates both the residual binding potential and the external laser field, and accounts for tunneling and quantum interference. 
With that purpose in mind, we will use the Coulomb Quantum-Orbit Strong-Field Approximation (CQSFA) \cite{Lai2015a}. The CQSFA is a path-integral strong-field approach that accounts for the driving field and the residual binding potential on equal footing, and has been applied by us to photoelectron holography by monochromatic fields \cite{Lai2017,Maxwell2017,Maxwell2017a,Maxwell2018,Maxwell2018b,Bray2021}. Apart from excellent agreement with experiments \cite{Maxwell2020,Kang2020,Werby2021}, the CQSFA allows unprecedented control about what type of interference leads to specific structures, as specific types of orbits may be switched on and off at will. 

This article is organized as follows. In Sec.~\ref{sec:background}, we briefly review the CQSFA and state the main assumptions used in this work. Subsequently, in Sec.~\ref{sec:symmetries}, we focus on the symmetries identified for linearly polarized bichromatic fields of commensurate frequencies, both by looking at the field and the saddle-point equations (Secs.~\ref{sub:fieldsymm} and \ref{sub:saddlesym}, respectively). Examples of how different symmetries affect the photoelectron momentum distributions are provided in Sec.~\ref{sec:Coulomb} for a specific case. Finally, in Sec.~\ref{sec:conclusions} we summarize the paper and state the main conclusions to be drawn from this work.

\section{Background}
\label{sec:background}
\subsection{General expressions}
We will focus on the transition amplitude $\left\langle\psi_{\mathbf{p}_f}(t) |U(t,t_0) |\psi_{0} \right\rangle $ from a bound state $\left\vert \psi _{0}\right\rangle $ to a final continuum state $ |\psi_{\mathbf{p}_f}(t)\rangle$ with momentum $\mathbf{p}_f$. The key difficulty is to calculate the time evolution operator, which, in atomic units, reads 
\begin{equation}
U(t,t_0)=\mathcal{T}\exp \bigg [i \int^t_{t_0}H(t^{\prime})dt^{\prime} \bigg],
\label{eq:Ufull}
\end{equation}
where $\mathcal{T}$ denotes time-ordering,
associated with the full Hamiltonian
\begin{equation}
    H(t)= H_a+H_I(t),
    \label{eq:1e-tdse}
 \end{equation}
where 
 \begin{equation}
H_a=\frac{\hat{\mathbf{p}}^{2}}{2}+V(\hat{\mathbf{r}})
\label{eq:Hatom}
\end{equation}
gives the field-free one-electron atomic Hamiltonian and $H_I(t)$ gives the coupling with the field. In equation~(\ref{eq:Hatom}),  $\hat{\mathbf{r}}$ and $\hat{\mathbf{p}}$ denote the position and momentum operators, respectively. Throughout, we employ the length gauge, so that $H_I(t)=\hat{\mathbf{r}}\cdot \mathbf{E}(t)$, use atomic units and consider a Hydrogen atom, so that  $V(\mathbf{r})=-1/|\mathbf{r}|$. The time-dependent Schr\"odinger equation can be either solved numerically, such as in \cite{qprop}, or used for constructing approximate, semi-analytic methods (for reviews see \cite{Symphony,Faria2020}).  Here, we will state the key assumptions employed in the Coulomb Quantum Orbit Strong-Field Approximation (CQSFA). As a benchmark, we use the freely available Schr\"odinger solver Qprop
 \cite{qprop,Mosert2016,Tulsky2020QPROP}. We specifically consider the version 3.2, which is essentially that in \cite{Tulsky2020QPROP} with a few modifications.

A convenient starting point is the integral equation 
\begin{equation}
U(t,t_0)=U_a(t,t_0)-i\int^t_{t_0}U(t,t^{\prime})H_I(t^{\prime})U_a(t^{\prime},t_0)dt^{\prime}\,
,\label{eq:Dyson}
\end{equation}
where $U_a(t,t_0)=\exp[iH_a (t-t_0)]$ is the time-evolution operator associated with the field-free Hamiltonian (\ref{eq:Hatom}). Using equation~(\ref{eq:Dyson}), one may write the transition amplitude from an initial bound state $|\psi _0(t^{\prime})\rangle$ to a final asymptotic state $|\psi_{\mathbf{p}_f}(t)\rangle$ as 
\begin{equation}
M(\mathbf{p}_f)=-i \lim_{t\rightarrow \infty} \int_{-\infty }^{t }d
t^{\prime}\left\langle \psi_{\mathbf{p}_f}(t)
|U(t,t^{\prime})H_I(t^{\prime})| \psi _0(t^{\prime})\right\rangle \,
,\label{eq:transitionampl}
\end{equation}
 with $\left | \psi _0(t^{\prime})\right\rangle=\exp[iI_pt']\left\vert \psi _{0}\right\rangle $, where $I_p$ is the ionization potential and $\mathbf{p}_f$ the final momentum. We assume that the electron is initially in the ground state, so that $I_p=0.5$ a.u. One should note that no approximation has been made in the time propagation described by equation~(\ref{eq:transitionampl}). 

Using path-integral methods and time-slicing techniques \cite{Kleinert2009,Milosevic2013JMP} in equation~(\ref{eq:transitionampl}), one obtains the expression 
\begin{eqnarray}\label{eq:CQSFATamp}
M(\mathbf{p}_f)&=&-i\lim_{t\rightarrow \infty
}\int_{-\infty}^{t}dt' \int d\mathbf{\tilde{p}}_0
\int_{\mathbf{\tilde{p}}_0}^{\mathbf{\tilde{p}}_f(t)} \mathcal {D}'
\mathbf{\tilde{p}}  \int
\frac{\mathcal {D}\mathbf{r}}{(2\pi)^3}  \nonumber \\
&& \times  e^{i S(\mathbf{\tilde{p}},\mathbf{r},t,t')}
\langle
\mathbf{\tilde{p}}_0 | H_I(t')| \psi _0  \rangle \, ,
\end{eqnarray}
where $\mathcal{D}'\tilde{\mathbf{p}}$ and $\mathcal{D}\mathbf{r}$ are the integration measures for the path integrals, and the prime indicates a restriction. These represent a sum over all possible paths in position and momentum, that the electron can take, between its start and end points. The tildes over the initial and intermediate momenta indicate field dressing, i.e., $\mathbf{\tilde{p}}_0=\mathbf{p}_0+\mathbf{A}(t')$ and $\mathbf{\tilde{p}}=\mathbf{p}+\mathbf{A}(\tau)$, with $t \le \tau \le t'$. 
This is the transition amplitude employed in the Coulomb Quantum-Orbit Strong-Field Approximation (CQSFA) \cite{Lai2015a,Lai2017,Maxwell2017}. One should note that the unitary transformation from the velocity to the length gauge leads to time-dependent momentum shifts that have been been incorporated in the intermediate momenta and in the matrix element from the bound state to the continuum. For the full derivation, together with the time-slicing techniques, see our previous publication \cite{Lai2015a}. 

The action in equation~(\ref{eq:CQSFATamp}) is given by
\begin{equation}\label{stilde}
S(\mathbf{\tilde{p}},\mathbf{r},t,t')=I_pt'-\int^{t}_{t'}[
\dot{\mathbf{p}}(\tau)\cdot \mathbf{r}(\tau)
+H(\mathbf{r}(\tau),\mathbf{p}(\tau),\tau)]d\tau,
\end{equation}
and the Hamiltonian by 
\begin{equation}
H(\mathbf{r}(\tau),\mathbf{p}(\tau),\tau)=\frac{1}{2}\left[\mathbf{p}(\tau)+\mathbf{A}(\tau)\right]^2
+V(\mathbf{r}(\tau)).
\label{Hamiltonianpath}
\end{equation}
One should note that, in equation~(\ref{eq:CQSFATamp}), both the residual binding potential and the external laser field are fully incorporated in the electron dynamics. Thus, in principle, the CQSFA considers rescattering in its full extent \cite{Lai2015a}. In the limit of vanishing Coulomb potential, equation~(\ref{eq:CQSFATamp}) reduces to the Strong-Field Approximation (SFA) transition amplitude associated with direct ATI\footnote{The direct ATI transition amplitude describes a process in which an electron leaves the atom, propagates in the continuum and reaches the detector without further interaction. It is obtained by replacing the full time-evolution operator by the Volkov time-evolution operator $U^{(V)}(t',t)$ in (\ref{eq:transitionampl}). The latter operator is associated with the Volkov Hamiltonian, in which the atomic binding potential is neglected. For reviews on this widespread approach, see, e.g., \cite{Ivanov2005,Popruzhenko2014a,Becker2018,Symphony} and for seminal papers see \cite{Lewenstein1995,Lohr1997,Becker1997}. In the context of direct strong-field ionization, the SFA is also known as the Keldysh-Faisal-Reiss (KFR) theory \cite{Keldysh1965,Faisal1973,Reiss1980}.}.

We solve the above-mentioned transition amplitude using a two-pronged contour in time. In the first half of the contour, the real part of the time is fixed and the imaginary part goes to zero. This means that the time integrals are performed from an initial complex time $t'=t'_r+it'_i$ to a a real time $t'_r$.
The second part of the contour is taken to be along the real time axis, up to infinity, that is, from $t'_r$ to a final time $t \rightarrow \infty$.  This choice of contour is widespread in approaches which incorporate the Coulomb potential \cite{Popruzhenko2008,Yan2012,Torlina2012,Torlina2013}. 

With such a contour choice, the action is written as
\begin{equation}\label{ssplit}
S(\mathbf{\tilde{p}},\mathbf{r},t,t')=S^{\mathrm{tun}}(\mathbf{\tilde{p}},\mathbf{r},t'_r,t')+S^{\mathrm{prop}}(\mathbf{\tilde{p}},\mathbf{r},t,t'_r),
\end{equation}
where 
$S^{\mathrm{tun}}(\mathbf{\tilde{p}},\mathbf{r},t'_r,t')$ and $S^{\mathrm{prop}}(\mathbf{\tilde{p}},\mathbf{r},t,t'_r)$ give the contribution of tunneling and continuum propagation to the action, respectively. The coordinate $\mathbf{r}(\tau)$ in the first part of the contour is commonly referred to as ``the tunnel trajectory". 

An approximation which makes finding the tunnel trajectory much easier is to assume the momentum remains constant along the first arm of the contour. This yields
\begin{equation}
\mathbf{r}_0(\tau) = \int^{\tau}_{t'}(\mathbf{p}_0+\mathbf{A}(t''))dt'',
\label{subBarrierPosition}
\end{equation}
so that $S^{\mathrm{tun}}(\mathbf{\tilde{p}},\mathbf{r},t'_r,t')$ is simplified to 
\begin{equation}\label{stun}
S^{\mathrm{tun}}(\mathbf{\tilde{p}},\mathbf{r},t'_r,t')=I_p(it'_i)-\int^{t'_r}_{t'}
H(\mathbf{r}_0(\tau),\mathbf{p}(t'_r),\tau)d\tau.
\end{equation}

Using saddle point methods to approximate the integrals in equation~(\ref{eq:CQSFATamp}) leads to the following system of saddle point equations:
\begin{eqnarray}
[\mathbf{p}(t')+\mathbf{A}(t')]^2 = -2I_p \label{SPEt}\\
\mathbf{\dot{r}}(\tau) = \mathbf{p}(\tau) + \mathbf{A}(\tau) \label{SPEp}\\
\mathbf{\dot{p}}(\tau) = -\nabla_rV(\mathbf{r}(\tau)). \label{SPEr}
\end{eqnarray}
Equation (\ref{SPEt}) governs the tunneling time $t'$ and has the form above due to the approximation that the sub-barrier momentum is constant. This approximation leads to the binding potential vanishing in the tunneling equation and has been discussed in detail in \cite{Maxwell2017}. Although, after these approximations, the CQSFA tunneling equation is mathematically identical to that in the SFA, the ionization times and the initial momenta will differ from their SFA counterparts as they must be matched at the tunnel exit with  the CQSFA results from the full Coulomb-distorted continuum propagation. For a discussion of practical implementations see \cite{Lai2015a}. Equations (\ref{SPEp}) and (\ref{SPEr}) determine the continuum trajectory of each orbit and it can be seen that the classical equations of motion have been recovered. 

Assuming that the tunnel exit is restricted to the laser's polarisation axis and setting it to be real, it can be approximately defined as
\begin{equation}
z_0 = Re[r_{0||}(t'_r)].
\label{tunnelExitGeneral}
\end{equation}
One should notice that this is an approximation, and that, in a more rigorous setting, complex equations of motion must be solved. Early studies for circularly polarized fields have shown that the imaginary parts of electron orbits in the continuum lead to electron deceleration \cite{Torlina2013}, in agreement with ab-initio computations \cite{Torlina2014}. For linearly polarized fields, complex orbits in the continuum will require dealing with branch cuts upon acts of rescattering \cite{Popruzhenko2014b,Pisanty2016}. For details in the context of the CQSFA  see \cite{Maxwell2018b,Faria2020}. Semiclassical methods from other research areas employing real orbits, such as the Herman Kluk propagator, will result in a dephasing for longer times due to the fact that tunneling is not properly incorporated (for a discussion in the strong-field context see our previous publication \cite{Zagoya2014}).

Within this approximation, the CQSFA transition amplitude (\ref{eq:CQSFATamp}) reads
\begin{equation}
\label{eq:MpPathSaddle}
M(\mathbf{p}_f)\propto-i \lim_{t\rightarrow \infty } \sum_{s}\bigg\{\det \bigg[  \frac{\partial\mathbf{p}_s(t)}{\partial \mathbf{r}_s(t'_s)} \bigg] \bigg\}^{-1/2} \hspace*{-0.6cm}
\mathcal{C}(t'_s) e^{i
	S(\mathbf{\tilde{p}}_s,\textbf{r}_s,t,t'_s)} 
\end{equation}
involving a sum over all orbits one would like to contribute to the final momentum distributions, 
where $t'_s$, $\mathbf{p}_s$ and $\mathbf{r}_s$ are the stationary variables obtained by solving the saddle-point equations. The term in brackets varies with the stability of the orbit while the term
\begin{equation}
\label{eq:Prefactor}
\mathcal{C}(t'_s)=\sqrt{\frac{2 \pi i}{\partial^{2}	S(\mathbf{\tilde{p}}_s,\textbf{r}_s,t,t'_s) / \partial t'^{2}_{s}}}\langle \mathbf{p}+\mathbf{A}(t'_s)|H_I(t'_s)|\Psi_{0}\rangle
\end{equation}
encodes the geometry of the initial electronic orbital, which, in the present publication, we take to be $1s$. For other types of orbitals in the CQSFA we refer to \cite{Kang2020,Maxwell2020,Werby2021,Bray2021}.

In practice, we employ the stability factor $\partial
\mathbf{p}_s(t)/\partial \mathbf{p}_s(t'_s)$ instead of that in equation~(\ref{eq:MpPathSaddle}), which may be obtained with a Legendre transformation. This choice will not influence the action if the electron starts from the origin \cite{Lai2015a}. Throughout, we will call the product of the stability factor with $\mathcal{C}(t'_s)$ ``the prefactor''. The CQSFA is solved as a boundary problem in which the initial conditions are written as functions of the final momenta, that is, given a final momentum $\mathbf{p}_f$ we seek an initial momentum $\mathbf{p}_0$ at the tunnel exit such that the saddle-point equations are satisfied. The final time $t$ is chosen to be at least 20 cycles long. More details about how the method is implemented can be found in our early publications \cite{Lai2015a,Lai2017,Maxwell2017}.

For a monochromatic field, the saddle-point solutions will lead to four types of orbits. An electron along orbit 1 leaves the atom and goes directly to the detector, without changing direction. In contrast, an electron along orbit 2 will be released half a cycle later or earlier, follow a field-dressed Kepler hyperbola and reach the detector without changing its momentum component perpendicular to the laser-field polarization. This behavior is similar to that of orbit 3, with the difference that, in the latter case, due to the residual potential the signs of the initial and final transverse momentum components will change. Finally, an electron along orbit 4 will be released on the same side as orbit 1, but will go around the core  before ultimately reaching the detector. These orbits have been first identified in \cite{Yan2010}, and have been discussed extensively in our previous publications. For bichromatic fields, this classification will change, but the relevant orbit types will depend on the field parameters. An example will be provided in Sec.~\ref{sec:Coulomb} for the situation in which the half-cycle symmetry is broken, but the high-frequency wave is weak.

Finally, an important practical issue is that, due to the necessity of taking a finite range of ionization times, there will be some arbitrariness about the initial and final times defining this range. This will lead to specific unit cells, which will influence the resulting holographic patterns depending on how they are chosen. Considering many cycles will overcome this arbitrariness, but a coherent sum will lead to strong ATI rings, which will obfuscate the remaining interference patterns. This is particularly critical if one is interested in intra-cycle interference.  An incoherent sum over unit cells has been employed in our previous publication \cite{Werby2021}, for a monochromatic field, in good agreement with experiments in which ATI rings are filtered out.

For a general polychromatic linearly polarized electric field
\begin{equation}
    E(t)=\sum_nE_nf_n( t),
\end{equation}
of amplitudes $E_n$ and time profiles $f_n( t)$, shifting the unit cell is equivalent to taking $f_n( t) \rightarrow f_n(t+t_{\mathrm{cell}})$ in the above equation. An incoherent sum over $t_{\mathrm{cell}}$ will eliminate this arbitrariness, and has been employed in our previous publication \cite{Werby2021} for a linearly polarized monochromatic field. 

\subsection{Linearly polarized bichromatic fields}

In the present work, we consider a two-color linearly polarized field composed of waves with commensurate frequencies $r\omega$ and $s\omega$, where $r,s$ are integers, phase difference $\phi$, and electric field amplitudes $E_r$, $E_s$. This gives an electric field of the form
\begin{equation}
\mathbf{E}_{r, s, \phi} (t)= [E_r\sin(r\omega t) + E_s\sin(s\omega t - \frac{s}{r}\phi)]\mathbf{e_{||}} ,   \label{eq:electric_field}
\end{equation}
and the vector potential of the form
\begin{equation}
    \mathbf{A}_{r, s, \phi} (t) = \left[\frac{E_r}{r\omega}\cos(r\omega t) + \frac{E_s}{s\omega}\cos(s\omega t - \frac{s}{r}\phi)\right]\mathbf{e_{||}}   
    \label{eq:vector_potential} 
\end{equation}
and the ponderomotive energy
\begin{equation}
U_p = \frac{E_r^2}{4 r^2 \omega^2} + \frac{E_s^2}{4 s^2\omega^2}.  \label{eq:ponderomotive_energy} 
\end{equation}
We will adopt the notation $(r,s)$ for the two frequencies involved, whereby the first and second index relates to the first and second wave, respectively \cite{Jasarevic2020,Habibovich2021} and refer to it as a $(r,s)$ field. 

Below we state the explicit expressions for the integral in the tunneling arm of the contour, the tunnel exit and the saddle-point equation associated with tunnel ionization. 

The action $S^{\mathrm{tun}}(\mathbf{\tilde{p}},\mathbf{r},t'_r,t')$  along the tunneling contour reads 
\begin{equation}\label{eq:stunintegrated}
\eqalign{S^{\mathrm{tun}}(\mathbf{\tilde{p}},\mathbf{r},t'_r,t') &= \left[I_p + U_p + \frac{1}{2}(p_{\parallel}(t'_r)^2+p_{\perp}(t'_r)^2) \right](it'_i)  \\
&- \frac{p_{\parallel}(t'_r) E_r}{(r\omega)^2}\left[\sin{(r\omega t)}\right]^{t'_r}_{t'} - \frac{p_{\parallel}(t'_r) E_s}{(s\omega)^2}\left[\sin{\left(s\omega t - \frac{s \phi}{r}\right)}\right]^{t'_r}_{t'} \\
&- \frac{E_r^2}{(2r\omega)^3} \left[\sin(2r\omega t)\right]^{t'_r}_{t'} - \frac{E_s^2}{(2s\omega)^3} \left[\sin{\left(2s\omega t - 2\frac{s \phi}{r}\right)}\right]^{t'_r}_{t'} \\
&- \frac{E_rE_s}{2rs\omega^3}\left[\frac{ \sin{\left(\omega t(r+s) - \frac{s\phi}{r}\right)}}{r+s}+\frac{ \sin{\left(\omega t(s-r) - \frac{s\phi}{r}\right)}}{s-r}\right]^{t'_r}_{t'}\\
&- \int^{t'_r}_{t'}V(\mathbf{r}_0(\tau))d\tau .}
\end{equation}
Equation (\ref{eq:stunintegrated}) is important for determining the saddle-point equation for the ionization times $t'$, which will be used to understand the symmetries governing the contrast and prominence of specific holographic patterns. The action $S^{\mathrm{prop}}(\mathbf{\tilde{p}},\mathbf{r},t,t'_r)$ in the second arm of the contour will influence the continuum propagation and the interference patterns of the photoelectron distributions, which, for the parameter range employed here will only play a secondary role. The tunnel exit (\ref{tunnelExitGeneral}) is given by
\begin{equation}
z_0 = \frac{E_r}{r^2\omega^2}\sin{(r\omega t'_r)}(1-\cosh{(r\omega t'_i)}) + \frac{E_s}{s^2\omega^2}\sin{\left(s\omega t'_r-\frac{s}{r}\phi\right)}(1-\cosh{(s\omega t'_i)}).
\label{tunnelExitBi}
\end{equation}

\section{Saddles and symmetries}
\label{sec:symmetries}
\subsection{Field symmetries}
\label{sub:fieldsymm}
We will investigate what symmetries may be present in the external field (\ref{eq:electric_field}) and the vector potential  (\ref{eq:vector_potential}). With that aim in mind, let us start from a general formulation and consider the field to be a periodic function in $t$ with period $T$. For simplicity, we will omit the unit vector $\mathbf{e}_{||}$ as this is essentially a one-dimensional problem. 

We define three operations on smooth functions, here denoted by $f(t)$. 
These can be time reflection around $\tau$ ($\mathcal{T}_R(\tau)$), a reflection with regard to the time axis ($\mathcal{F}$), and time translation by $\tau $ ($\mathcal{T}_T(\tau)$) such that
\begin{eqnarray}
\centering
\mathcal{T}_R(\tau) f(\tau + t) &=& f(\tau - t), \label{eq: Time reversal} \\
\mathcal{T}_T(\tau) f(t) &=& f(t - \tau), \label{eq: Time Translation} \\ 
\mathcal{F} f(t) &=& -f(t), \label{eq: Force reversal} \
\end{eqnarray}
for real times, $t$.  If  $E_{r, s, \phi}(t)$ and $A_{r, s, \phi}(t)$  have period $T$, this is equivalent to saying that $E_{r, s, \phi}(t)$ and $A_{r, s, \phi}(t)$ have $\mathcal{T}_T(T)$ symmetry, which is broken by taking a short pulse of light, but not by introducing a second colour.

\begin{figure}
    \centering
    \includegraphics{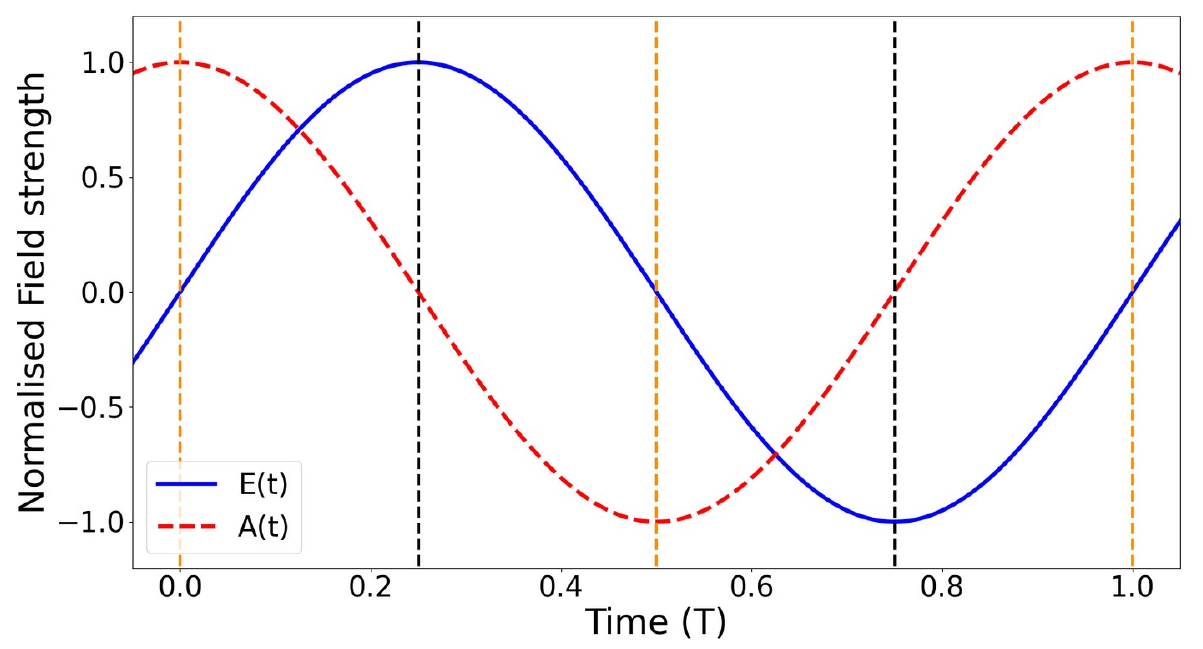}
    \caption{Time profile of a linearly polarized monochromatic field and the corresponding vector potential, which exhibit the three symmetries (i) - (iii). Black lines at $(2n+1)T/4$ indicate the times that permit $\mathcal{T}_R$ symmetry for the electric field, while orange (light grey) lines at $nT/2$ indicate the times that permit $\mathcal{F}\mathcal{T}_R$ symmetry for the electric field. For the vector potential, these times are interchanged}
    \label{fig:symmsmono}
\end{figure}

There are three symmetries for monochromatic linearly polarized fields, which can be broken by introducing a second colour. A monochromatic field remains invariant under: 
\begin{enumerate}
    \item  a translation of half a cycle followed by a reflection with regard to the time axis, that is, $\mathcal{F}\mathcal{T}_T\left(\frac{T}{2}\right)E(t)=E(t)$. This is known as the half-cycle symmetry, and usually written as $E(t\pm T/2)=-E(t)$.
    \item a time reflection around its extrema, so that $\mathcal{T}_R\left(\tau_{ex}\right)E(t)=E(t)$, where $\tau_{ex}$ are the times for which the extrema occur.
    \item a time reflection around its zero crossings followed by a reflection with regard to the time axis, that is, $\mathcal{F}\mathcal{T}_R\left(\tau_{cr}\right)E(t)=E(t)$, where, similarly $\tau_{cr}$ are the times for which the zero crossings take place.
\end{enumerate}
All these properties hold for the electric field and the vector potential, but with  $\tau_{cr}$ and  $\tau_{ex}$ swapped. An example is provided  in Table \ref{table: monoSym} and illustrated in Fig.~\ref{fig:symmsmono} for a sine field of frequency $\omega$.
 \begin{table}[h]
 \begin{center}
 \begin{tabular}{||c | c | c||} 
  \hline
  {} & $E^{\mathrm{mono}}(t)$ & $A^{\mathrm{mono}}(t)$  \\ [1ex] 
 
 Half-cycle symmetry & $\mathcal{F}\mathcal{T}_T\left(\frac{T}{2}\right)$ & $\mathcal{F}\mathcal{T}_T\left(\frac{T}{2}\right)$  \\ [2ex] 

  Reflection around extrema & $\mathcal{T}_R\left(\frac{(2n+1)T}{4}\right)$ & $\mathcal{T}_R\left(\frac{nT}{2}\right)$ \\ [2ex] 

 Reflection around zero crossings & $\mathcal{F}\mathcal{T}_R\left(\frac{nT}{2}\right)$ & $\mathcal{F}\mathcal{T}_R\left(\frac{(2n+1)T}{4}\right)$  \\ [1ex] 
  \hline
\end{tabular}
\end{center}
\setlength{\abovecaptionskip}{0pt plus 3pt minus 2pt}
\caption{Symmetries satisfied by a sinusoidal monochromatic linearly polarized electric field of frequency $\omega$. The first column specifies the symmetry, while the remaining columns provide information about the relevant times for the electric field and the vector potential, respectively. Throughout, $n$ is an integer.}
 \label{table: monoSym}
 \end{table}

For a two colour field, the above symmetries may be retained or broken. The set of values of $\phi$ for which the field retains the symmetry defined by the symmetry operation $\mathcal{O}(\tau)$ can be written as
\begin{equation}
\label{eq: Symmetric phis}
\Phi_{r, s}(\mathcal{O}(\tau)) = \{\phi | E_{r, s, \phi} = \mathcal{O}(\tau)E_{r, s, \phi}\}.
\end{equation}

We can use this to write statements about which of the monochromatic symmetries are retained when a second colour is added. 
For instance, for real times, $\tau_1$ and $\tau_2$
\begin{eqnarray}
\centering
\phi \in \Phi_{r,s}(\mathcal{F}\mathcal{T}_R(\tau_1)) \cap \Phi_{r,s}(\mathcal{T}_R(\tau_2)) \Rightarrow \phi \in \Phi_{r,s}(\mathcal{F}\mathcal{T}_T(T/2)), \label{eq: Symmetry relationship} \
\end{eqnarray}
which states that if the field has symmetry under time reflection around its extrema as well as its zero crossings then the half cycle symmetry must also exist for this field. One should note that if the half-cycle symmetry holds, this is no guarantee that the other two are present.

For an $(r,s)$ field,  all combinations of $(r, s)$ can be reduced either to the case in which $r$ and $s$ have opposite parity or to both $r$ and $s$ being odd.  One should note that if  $(r, s)$ are not coprimes, with regard to symmetry it will reduce to one of these cases scaled by the multiplicative factor that transformed the indices.

For $r$ and $s$ both odd, the half cycle symmetry is preserved so for a given $\phi$ either both of the other time reflection symmetries hold or neither do.
For $r$ and $s$ with opposite parity, the half cycle symmetry is broken, so only one of the other time reflection symmetries may be satisfied for a given $\phi$.
This is because the statement (\ref{eq: Symmetry relationship}) is true  if one interchanges the operations. This is due to the structure of the symmetry group. The symmetry group of the temporal evolution of the monochromatic field has only 4 subgroups which describe the symmetries of periodic mathematical objects. These correspond precisely to the objects described: the periodic field with no further symmetry, the field with just time reflection, the field where time reflection combined with reflection along the time axis is a symmetry, and the field with just half cycle symmetry. The symmetry groups here are simply 5 of the 7 Frieze groups \cite{Gallian}.

For a given $\phi$, if the field is symmetric with regard to $\mathcal{F}\mathcal{T}_R(\tau)$ or $\mathcal{T}_R(\tau)$, then this symmetry will also hold for $\mathcal{F}\mathcal{T}_R(\tau+nT/2)$ and $\mathcal{T}_R(\tau+nT/2)$ respectively where $n\in\mathbb{Z}$. Thus, it makes sense to view symmetric points separated by $nT/2$ as equivalent so we only need to consider symmetries which exist within a half-cycle of the field. This is not to be mistaken with the half cycle symmetry.

\begin{figure}[h!]
\centering
\includegraphics[width=\textwidth]{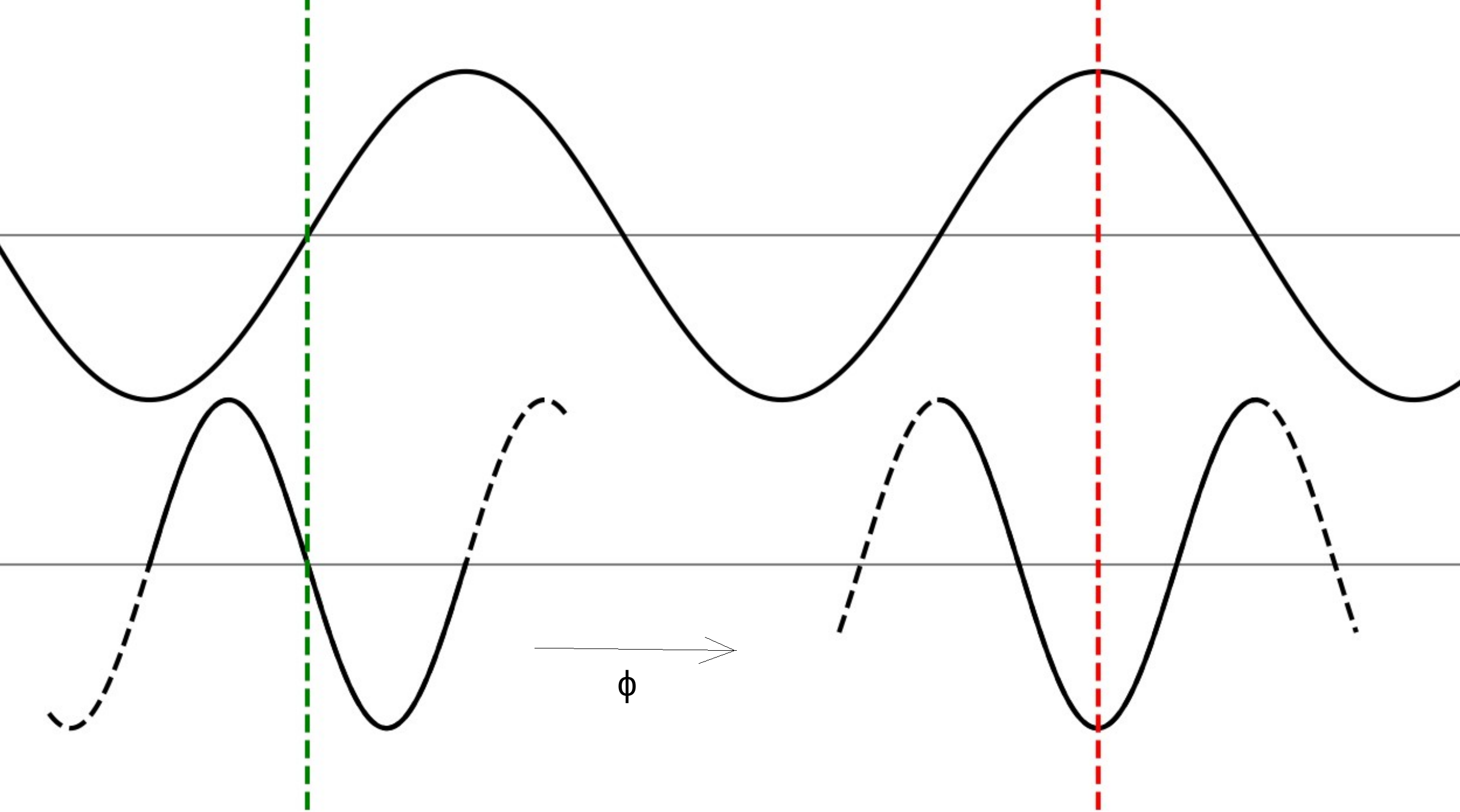}
\caption{\small This schematic shows the ways in which the two additive components of the bichromatic field must be aligned to permit a symmetry. By varying the phase $\phi$, the alignment of these constituent fields can be varied. The green dashed line indicates that fields where the two constituent fields have zeroes at the same time will have symmetric gradient around this field zero crossing. The red dashed line indicates that fields where the constituent fields have extrema which coincide will have symmetry around this extreme point.}
\label{fig:peak_matching}
\end{figure}
The symmetries $\mathcal{F}\mathcal{T}_R(\tau)$  and $\mathcal{T}_R(\tau)$ for two-color fields may be understood in terms of overlapping field maxima or zero crossings. For a field as defined by equation (\ref{eq:electric_field}), all possible combinations of field zero crossings coinciding as in Fig.~\ref{fig:peak_matching} can be enumerated. This gives the condition that for
\begin{equation}
\centering
\phi_{\textrm{s}} = \frac{2\pi}{2s}(ns-mr), \label{eq:phases symmetric under rotation appendix} \
\end{equation}
where $n,m \in \mathbb{Z}$, there exists some $\tau$ such that the $\mathcal{F}\mathcal{T}_R(\tau)$ symmetry holds.

Likewise, the same can be done for the phases $\phi$ such that the field extrema coincide. For
\begin{equation}
\centering
\phi_{\textrm{s}}= \frac{2\pi}{4s}((2n+1)s-(2m+1)r), \label{eq:phases symmetric under reflection appendix} \
\end{equation}
where $n,m \in \mathbb{Z}$, there exists $\tau$ such that the $\mathcal{T}_R(\tau)$ symmetry holds.

These conditions can be simplified to equations 
\begin{equation}
\label{eq: phases symmetric under rotation}
\phi = \frac{q\pi}{2s} \textrm{ where } q \in 2\mathbb{Z},
\end{equation}
and 
\begin{equation}
\label{eq: phases symmetric under reflection}
\phi = \frac{k\pi}{2s} \textrm{ where }  k \in \left\{ \begin{array}{ll} 2\mathbb{Z} \quad\quad \textrm{    for $r$ and $s$ both odd} \\ 2\mathbb{Z}+1 \quad \textrm{for $r$ and $s$ of opposite parity}  \end{array} \right.,
\end{equation}
respectively. The latter formulations consider the cases for which $(r,s)$ are both odd or of opposite parity separately where necessary. 
The same reasoning in Fig.~\ref{fig:peak_matching} can be generalized for N-colour fields so the conditions for each of the relative phases to give a field with a certain symmetry can be found in the same way.

\begin{figure}
    \centering
    \includegraphics[width=\textwidth]{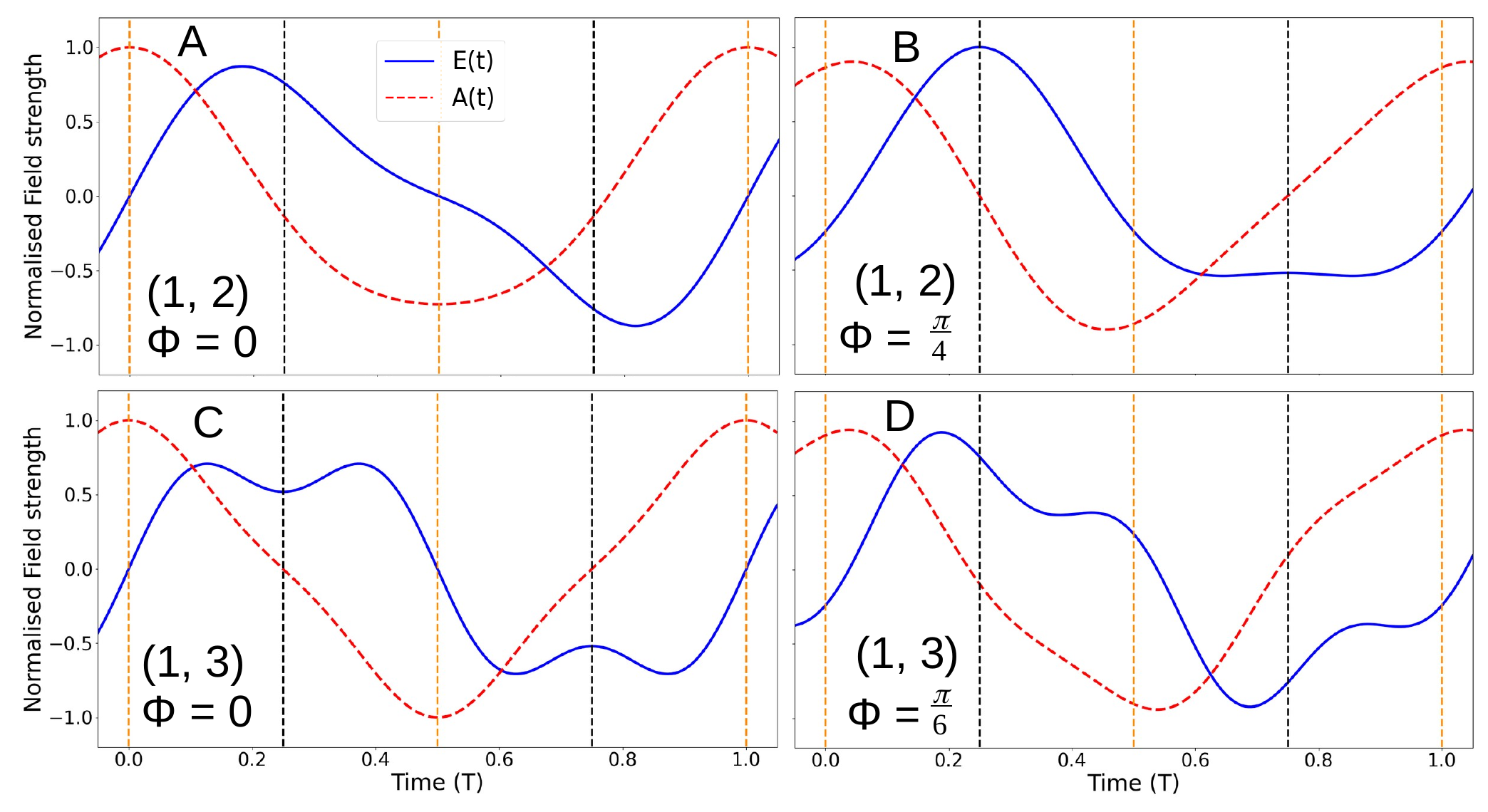}
    \caption{The electric field is shown by the blue (solid grey) line and the vector potential by a red (dashed grey) line. In panels A and B bi-chromatic fields with frequency (1, 2) are plotted while in panels C and D, bi-chromatic fields with frequency (1, 3) are shown.   The vertical dotted lines represent the possible lines of symmetry. Black lines at $(2n+1)T/(4r)$ represent the times that permit $\mathcal{T}_R$ symmetry for the electric field, while orange (light grey) lines at $nT/(2r)$ represent the times that permit $\mathcal{F}\mathcal{T}_R$ symmetry for the electric field. Varying the phase $\phi$ can change around which axis the field is symmetric and whether it is the electric field  or the gradient of the electric field which is symmetric. For clarity here the intensity ratio $I_r/I_s = 10$ while in the remainder of the paper much weaker second fields are considered.}
    \label{fig:fieldsymmetry}
\end{figure}

Fig.~\ref{fig:fieldsymmetry} illustrates all of the possible symmetry configurations (except a field with no symmetry) by showing a cycle of the electric field and vector potential for a selection of the parameters $r$, $s$ and $\phi$. The top two panels [Fig.~\ref{fig:fieldsymmetry}(A and B)] are for $(1, 2)$ fields. From equation (\ref{eq: phases symmetric under rotation}), taking $q=0$ means that the $\phi = 0$ field [Fig.~\ref{fig:fieldsymmetry}(A)] has $\mathcal{F}\mathcal{T}_R$ symmetry. If we consider equation (\ref{eq: phases symmetric under reflection}) and let $k=1$, then the $\phi = \pi/4$ field [Fig.~\ref{fig:fieldsymmetry}(B)] has $\mathcal{T}_R$ symmetry. The lower two panels [Figs.~\ref{fig:fieldsymmetry}(C) and (D)] are for $(1, 3)$ fields. Therefore, $q$ and $k$ from equations (\ref{eq: phases symmetric under rotation}) and (\ref{eq: phases symmetric under reflection}), respectively, must be even. This means that, for $\phi$ which are multiples of $\pi/3$ as in Fig.~\ref{fig:fieldsymmetry}(C), there exists both $\mathcal{F}\mathcal{T}_R$ and $\mathcal{T}_R$ symmetries so this field has all of the symmetries of the monochromatic field as described in Table \ref{table: monoSym}. For other values of $\phi$ [Fig.~\ref{fig:fieldsymmetry}(D)], only the half-cycle symmetry holds. Throughout Fig.~\ref{fig:fieldsymmetry}, $r=1$, which means that the times acting as the symmetry axis in each case coincide precisely with those in Table \ref{table: monoSym}. For some combinations $(r, s)$ where $r>1$, the axis about which the field is symmetric can vary non-trivially with the relative phase $\phi$ as is discussed in section \ref{sub:phasediagrams}.
In \ref{sec: fieldapp}, we propose parameters to quantify the degrees of asymmetry in a two-color field using the relative phase. 

\subsection{Saddle-point equations}
\label{sub:saddlesym}

In section \ref{sub:fieldsymm}, the real time symmetries of the electric field $E_{r,s,\phi}(t)$ (\ref{eq:electric_field}) have been investigated. However, the saddle point time $t'$ is complex, as tunneling is not a classically allowed process. The imaginary part Im[$t'$] is particularly relevant as it is an indicator of the importance of the orbit which tunnels at $t'$. This is because the most significant term in equation (\ref{eq:stunintegrated}) is the first, which is linear in Im[$t'$]. In the case of two interfering orbits $i$ and $j$, $\Delta\textrm{Im}[t'_{ij}]=|\textrm{Im}[t'_i] - \textrm{Im}[t'_j]|$ will indicate the level of contrast to be expected in the interference pattern. $\Delta$Im[$t'_{ij}$] $= 0$ implies maximum contrast as orbits $i$ and $j$ are equally relevant. Therefore, not only are there symmetries in $\mathrm{Re}(t')$, but also symmetries in $\mathrm{Im}(t')$ which are arguably more important. This can be investigated by looking at the symmetries of the saddle point times from equation (\ref{SPEt}). 

In this section, we will focus on the symmetries caused by the field only. Therefore, in the results that follow we display complex ionization times computed within the SFA. A proper treatment of the saddles in the CQSFA framework is non-trivial and has been done approximately in \cite{Maxwell2018b}. A more rigorous treatment is work in progress. Nonetheless, under the present approximations the CQSFA equation describing tunnel ionization is formally identical to its SFA counterpart, so that it can be used for an approximate study of the existing symmetries. One should note, for the CQSFA, the initial momenta and ionization times will differ due to the influence of the Coulomb potential; see, e.g., \cite{Lai2015a}.
In the plots that follow, namely Figs.~\ref{fig:momentum_range} and \ref{fig:ret_v_imt}, we will present the saddles for the upper half plane of $t'$, as they are physically significant \cite{Pisanty2016,Maxwell2018b}.

\subsubsection{Complex ionization times}
To establish the saddle point symmetries and to determine for which $\phi$ they occur, it helps to write saddle point equation (\ref{SPEt}) for the general bichromatic electric field (\ref{eq:electric_field}). This can be shown in a similar manner to \cite{Jasarevic2020} to be equation (\ref{eq: general two colour spe}) where the variable substitution $x = e^{i\omega t'}$ has been made.

\begin{equation}
\centering
\frac{E_s}{s}\left[\frac{x^s}{e^{i\phi s/r}}+\frac{e^{i\phi s/r}}{x^s}\right] + \frac{E_r}{r}\left[x^r+\frac{1}{x^r}\right] - 2\omega\left[p_{0\parallel} \pm i\sqrt{2I_p+p_{0\perp}^2}\right] = 0\label{eq: general two colour spe} \
\end{equation}

The phase $\phi=n\pi/(2s)$ for $n \in \mathbb{Z}$ covers all phases which admit a symmetry in equations (\ref{eq: phases symmetric under rotation}) and (\ref{eq: phases symmetric under reflection}) when $r$ and $s$ have opposite parity. For $(r, s)$ with opposite parity and such that $r$ and $s$ are co-prime, the following symmetries exist for saddle point times such that, for each $\mu \in [2s]$\footnote{The notation $[2s]$ here represents the set of integers between 1 and $2s$. It is just a labelling system for the $2s$ distinct saddles so any set with cardinality $2s$ will be sufficient.}, there is a $\nu \in [2s]$ 
\begin{eqnarray}
\centering
\mathrm{Re} [t'_\mu(p_{0\parallel})] &=& \frac{nT}{2}\left[r_s^{-1} + C\left(\frac{1}{r}-1\right)\right]-\mathrm{Re} [t'_\nu((-1)^n p_{0\parallel})] \quad \textrm{and} \label{eq: odd even real time solution transformation simple} \\
\mathrm{Im} [t'_\mu(p_{0\parallel})] &=& \mathrm{Im} [t'_\nu((-1)^n p_{0\parallel})].\label{eq: odd even imag time solution transformation simple} \
\end{eqnarray}
Here, 
\begin{equation}
 C= \frac{1-r_s^{-1}r}{s},
 \label{eq:C}
\end{equation}
with $r_s^{-1}$ representing the multiplicative inverse of $r$ modulo $s$.
It was specified before that $r$ and $s$ were co-prime as this ensures that such a multiplicative inverse actually exists.

Conversely, for $(r, s)$ which are both odd we additionally have the half cycle symmetry for all $\phi$. In this case, for $\mu \in [2s]$, there is $\nu \in [2s]$ so that the saddle point times have the symmetries
\begin{eqnarray}
\centering
\mathrm{Re} [t'_\mu(p_{0\parallel})] &=& \frac{T}{2} + \mathrm{Re} [t'_\nu(-p_{0\parallel})] \quad \textrm{and} \label{eq: odd half cycle real time solution transformation} \\
\mathrm{Im} [t'_\mu(p_{0\parallel})] &=& \mathrm{Im} [t'_\nu(-p_{0\parallel})].\label{eq: odd half cycle imag time solution transformation} \
\end{eqnarray}

Additionally, for $\phi = {n\pi}/{2s}$ such that $n \in 2\mathbb{Z}$, each $\mu \in [2s]$ can be paired with another $\nu \in [2s]$ so that the symmetries 
\begin{eqnarray}
\centering
\mathrm{Re} [t'_\mu(p_{0\parallel})] &=& \frac{nT}{2}\left[\frac{C}{r} + \frac{1}{2} \right]-\mathrm{Re} [t'_\nu((-1)^{n/2} p_{0\parallel})] \quad \textrm{and} \label{eq: odd real time solution transformation simple} \\
\mathrm{Im} [t'_\mu(p_{0\parallel})] &=& \mathrm{Im} [t'_\nu((-1)^{n/2} p_{0\parallel})]\label{eq: odd imag time solution transformation simple} \
\end{eqnarray}
hold. The combination of symmetries in equations (\ref{eq: odd half cycle real time solution transformation}), (\ref{eq: odd half cycle imag time solution transformation}), (\ref{eq: odd real time solution transformation simple}) and (\ref{eq: odd imag time solution transformation simple}) can be used to show that two further symmetries exist. These are
\begin{eqnarray}
\centering
\mathrm{Re} [t'_\mu(p_{0\parallel})] &=& \frac{nT}{2}\left[\frac{C}{r} + \frac{1}{2} \right] - \frac{T}{2} -\mathrm{Re} [t'_\nu(-(-1)^{n/2} p_{0\parallel})] \quad \textrm{and} \label{eq: odd real time solution transformation simple 2} \\
\mathrm{Im} [t'_\mu(p_{0\parallel})] &=& \mathrm{Im} [t'_\nu(-(-1)^{n/2} p_{0\parallel})].\label{eq: odd imag time solution transformation simple 2} \
\end{eqnarray}
Further details and the derivation of these symmetries can be found in \ref{sec: sadapp}. 

\begin{figure}[]
\centering
\includegraphics[width=\textwidth]{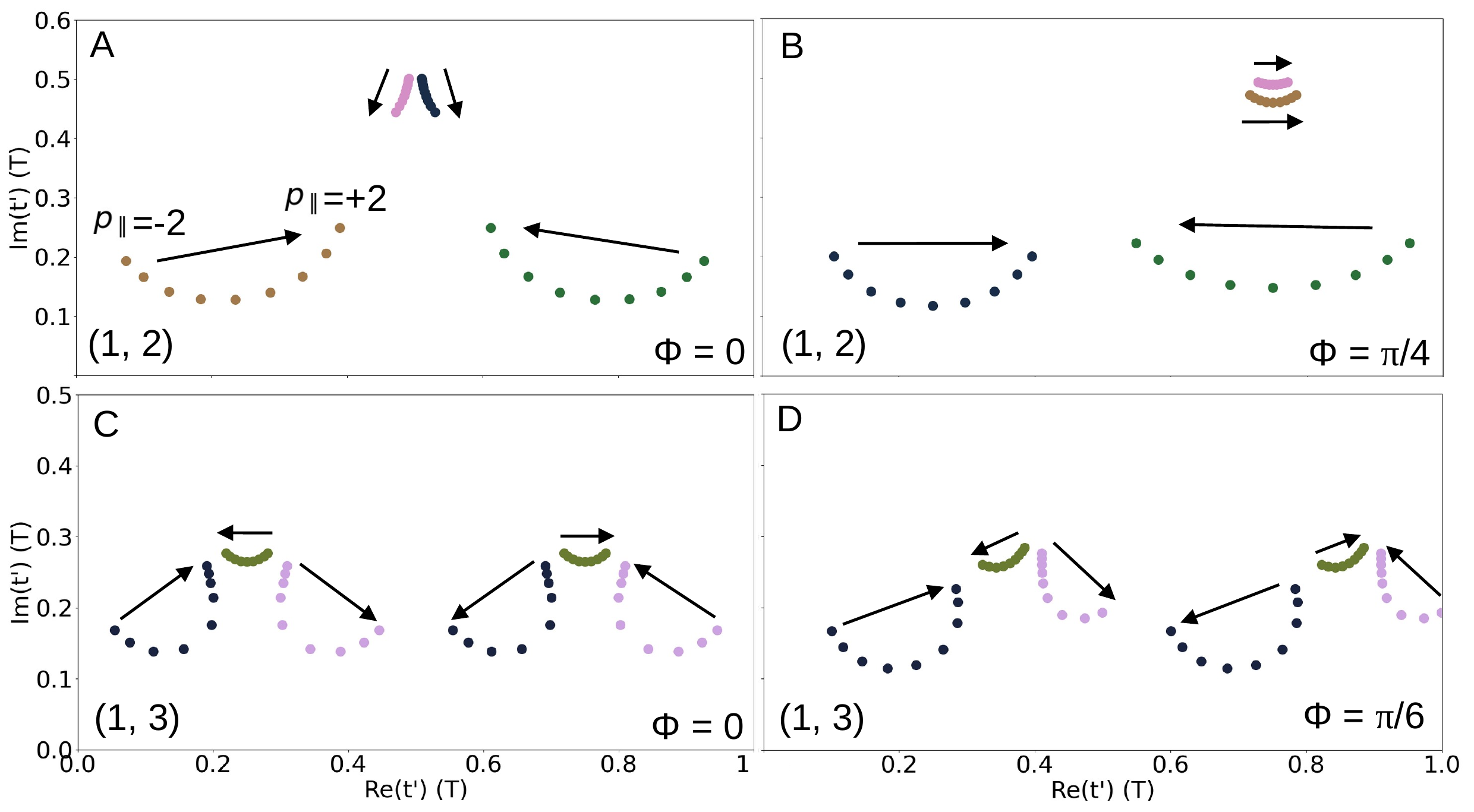}
\caption{\small The saddle point times, $t'$, as determined by equation (\ref{SPEt}), are plotted for nine distinct parallel momenta, uniformly distributed between $p_{0\parallel}=-2$ and $p_{0\parallel}=+2$ atomic units (increments of 0.5 atomic units), while the perpendicular component is kept constant at $p_{0\perp}=0.1$ atomic units. The black arrows drawn on the figure illustrate the direction that the saddle points will move as the parallel component of momentum is increased. It would be unclear if every saddle was labelled in terms of $p_{0\parallel}$ as has been done for a pair of saddles in panel (A). Therefore, the arrow head indicates which end of the sequence of scatter points is the saddle for $p_{0\parallel}=+2$ atomic units.  The four panels show the same selection of field type $(r,s)$ and phase $\phi$ as in Fig.~\ref{fig:fieldsymmetry}, so as to directly compare how the field symmetry affects the saddle symmetry. However, the intensity ratio is ${I_r}/{I_s}=100$ for panels (A) and (B) so that it matches the physical example which is explored in section \ref{sec:Coulomb}. Panels (C) and (D) use ${I_r}/{I_s}=50$ instead, for the sake of saddle distinguishability.}
\label{fig:momentum_range}
\end{figure}

\begin{figure}[]
\centering
\includegraphics[width=\textwidth]{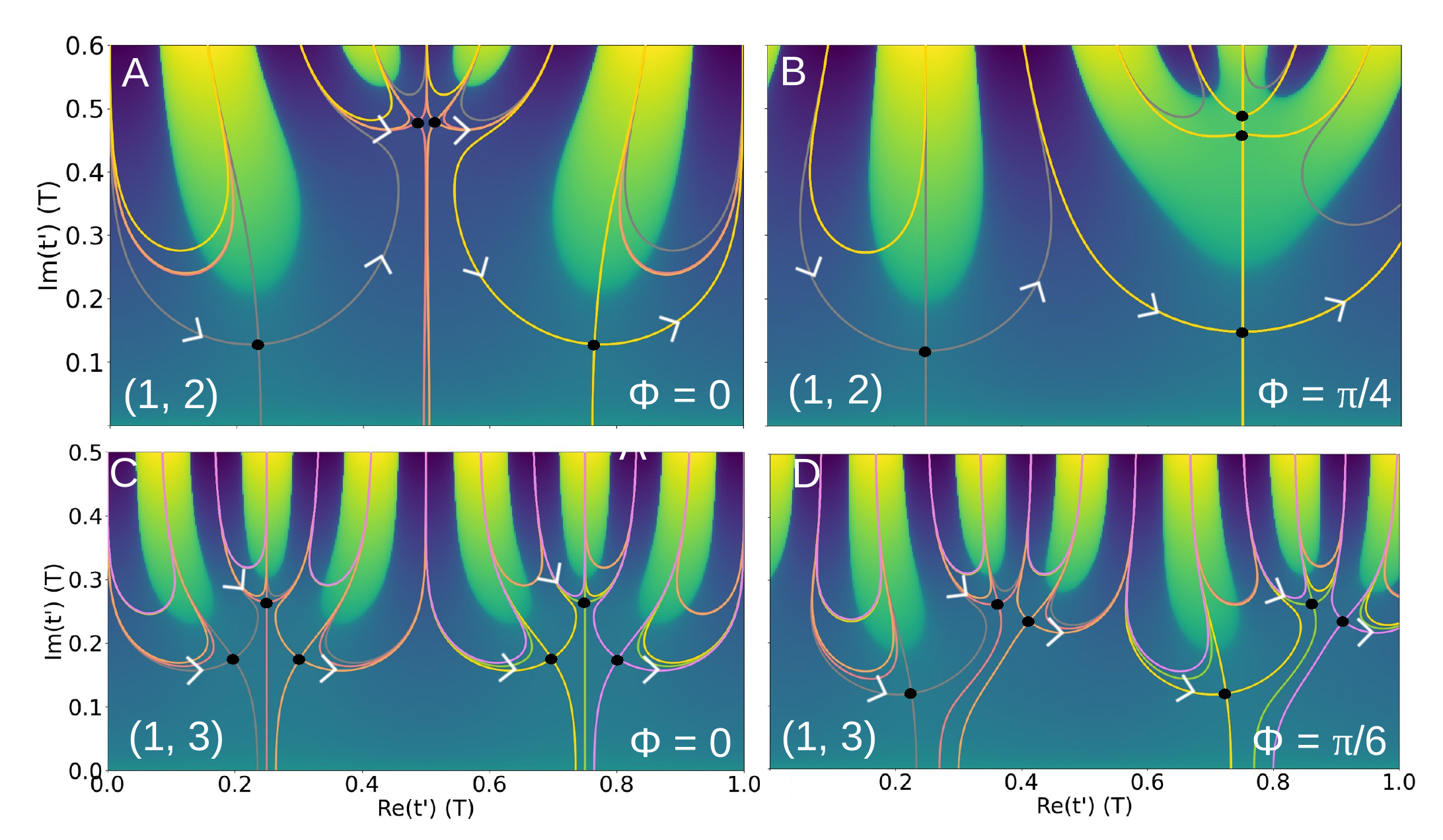}
\caption{\small Ionization times in the complex plane, together with steepest descent/ascent contours, for a model atom in several linearly polarized bichromatic fields, within the SFA framework. The sub barrier semi-classical action (\ref{eq:stunintegrated}) is shown by the colourmap scaled by the function $\sinh{(S)}$. The black dots mark the saddle point times calculated from equation \ref{SPEt} for $p_{0\parallel} = 0$ and $p_{0\perp} = 0.1$ atomic units. The coloured lines through the saddles are the contours of steepest descent/ascent and the white arrows indicate the direction of integration along the contours to be used for the method of steepest descent.  The panels (A), (B), (C) and (D) are calculated using the same field parameters as the respective panels in Fig.~\ref{fig:ret_v_imt}. }
\label{fig:ret_v_imt}
\end{figure}

For the field of type (1, 2) we derive the following very simple symmetries from equations (\ref{eq: odd even real time solution transformation simple}) and (\ref{eq: odd even imag time solution transformation simple}). For (1, 2), $r_s^{-1}=1$ so $C=0$ and this gives the transformations
\begin{eqnarray}
\centering
\mathrm{Re} [t'_\mu(p_{0\parallel})] &=& \frac{Tn}{2}-\mathrm{Re} [t'_\nu((-1)^n p_{0\parallel})] \quad \textrm{and}\label{eq: 1 2 real time solution transformation} \\
\mathrm{Im} [t'_\mu(p_{0\parallel})] &=& \mathrm{Im} [t'_\nu((-1)^n p_{0\parallel})]\label{eq: 1 2 imag time solution transformation} \
\end{eqnarray}
for phases which are of the form $\phi = n\pi/4$.
The only distinct cases are for $n$ odd and $n$ even. For $n$ even, $\phi \in \{..., -\pi, -\frac{\pi}{2}, 0, \frac{\pi}{2}, \pi, ...\}$ 
\begin{eqnarray}
\centering
\mathrm{Re} [t'_\mu(p_{0\parallel})] &=& T-\mathrm{Re} [t'_\nu(p_{0\parallel})] \quad \textrm{and} \label{eq: 1 2 n even real time solution transformation} \\
\mathrm{Im} [t'_\mu(p_{0\parallel})] &=& \mathrm{Im} [t'_\nu(p_{0\parallel})]\label{eq: 1 2 n even imag time solution transformation} \
\end{eqnarray}
This can be seen to hold in Fig.~\ref{fig:momentum_range}(A) by looking at the saddle point times for a range $p_{0\parallel}$. The $\mathcal{F}\mathcal{T}_R({T}/{2})$ symmetry leads to reflection symmetry of the saddles along the axis $\mathrm{Re} [t'] = {T}/{2}$. In particular, the beige (light grey) saddles are the mirror image of the green (dark grey) saddles while the pink (lightest grey) saddles are the mirror image of the dark blue (darkest grey) saddles.

For $n$ odd, $\phi \in \{..., -\frac{3\pi}{4}, -\frac{\pi}{4}, \frac{\pi}{4}, \frac{3\pi}{4}, ...\}$
\begin{eqnarray}
\centering
\mathrm{Re} [t'_\mu(p_{0\parallel})] &=& \frac{T}{2}-\mathrm{Re} [t'_\nu(-p_{0\parallel})] \quad \textrm{and} \label{eq: 1 2 odd real time solution transformation} \\
\mathrm{Im} [t'_\mu(p_{0\parallel})] &=& \mathrm{Im} [t'_\nu(-p_{0\parallel})]\label{eq: 1 2 odd imag time solution transformation} \
\end{eqnarray}
This is visible in Fig.~\ref{fig:momentum_range}(B) as the $\mathcal{T}_R({T}/{4})$ symmetry leads to saddles with reflection symmetry along the axis $\mathrm{Re} [t'] = {T}/{4}$ when combined with a momentum transformation $p_{0\parallel} \rightarrow -p_{0\parallel}$. The momentum transformation is necessary as the reflection reverses the direction of the black arrows in Fig.~\ref{fig:momentum_range}(B). It should be noted that for the field parameters used in Fig.~\ref{fig:momentum_range}(B), the equations (\ref{eq: 1 2 odd real time solution transformation}) and (\ref{eq: 1 2 odd imag time solution transformation}) are satisfied by setting $\mu=\nu$. This means that each saddle has the same colour as its mirror image. However, this will not be the case for arbitrary choice of the field parameters $E_r$ and $E_s$. This is because the steepest descent contours shown in Fig.~\ref{fig:ret_v_imt} undergo changes in topology as we move through the space of parameters, which can be associated to changes to the symmetry relations of the saddle points. A good example can be discussed by looking at Fig.~\ref{fig:momentum_range}(C) and Fig.~\ref{fig:ret_v_imt}(C). In this case, the field has all the same symmetries as the monochromatic field. Despite this, the saddles do not satisfy the same symmetry relations as the monochromatic saddle points do. The equations are of the same form, (\ref{eq: odd real time solution transformation simple}) and (\ref{eq: odd imag time solution transformation simple}), but their indices $\mu$ and $\nu$ differ. 
For a monochromatic field, equation~(\ref{eq: odd imag time solution transformation simple}), for the saddles' symmetry with regard to the axis $\mathrm{Re} [t'] = {T}/{4}$ is satisfied when $\mu = \nu$. This is visibly not the case for every set of saddles in Fig.~\ref{fig:momentum_range}(C). For instance, the dark blue (darkest grey) saddles do not have reflection symmetry around the axis $\mathrm{Re} [t'] = {T}/{4}$. Instead, the mirror image of the dark blue (darkest grey) saddles around $\mathrm{Re} [t'] = {T}/{4}$ is the pink (lightest grey) saddles with the additional momentum inversion as before. This means that if we label the dark blue saddles $t'_1$, the green saddles $t'_2$, and the pink saddles $t'_3$, then equation~(\ref{eq: odd imag time solution transformation simple}) is satisfied for the pair $\mu = 1$, $\nu = 3$ and for $\mu = \nu = 2$.  
The symmetry relation differs because the steepest descent contour topology shown in Fig.~\ref{fig:ret_v_imt}(C) is different to that of the monochromatic case. As $I_r/I_s$ is increased, the topology of the contours in Fig.~\ref{fig:ret_v_imt}(C) will undergo a transformation such that the symmetry relation of the saddles becomes identical to the monochromatic case ($\mu = \nu $ for all saddles). In this case saddles $t'_1(0)$ and $t'_3(0)$ will merge at a critical value of $I_r/I_s$ and beyond this point will satisfy equation~(\ref{eq: odd imag time solution transformation simple}), with $\mu = \nu = 1$ and $\mu = \nu = 3$.
In the remaining panel, Fig.~\ref{fig:momentum_range}(D), the half cycle symmetry of the field means that the saddle points repeat identically every half cycle. Therefore, there will be no resulting asymmetry in the final momentum distributions of tunneled electrons \cite{Habibovich2021}. Having the other symmetries broken means that the sub half cycle saddles have no symmetry, which can lead to changes to the contrast of holographic interference patterns.      

By looking at how the saddle points differ between the panels in Fig.~\ref{fig:ret_v_imt} it becomes clear why there is difficulty in devising a fully general orbit classification for arbitrary bichromatic fields. 
In panel (A), for $\phi = 0$, the reflection symmetry across $\mathrm{Re} [t'] = {T}/{2}$ means that we must have an identical number of saddles in each half cycle and each saddle is paired with another in the next half cycle. When the symmetry has been broken as in panel (B), for $\phi = \pi/4$, the saddles are unpaired and there can be a different number of saddles in each half cycle. In this case, there are now three saddles in one half cycle and just one in the other. This difference alone means that a general orbit classification must depend on $\phi$, as the half cycle in which an orbit begins in will significantly alter its behaviour. Another parameter to be considered when classifying orbits is the ratio $E_r/E_s$. In Figs.~\ref{fig:ret_v_imt}(A) and (B), there are a total of $2s = 4$ saddles in the upper half plane per field cycle. However, because of the large value of $E_1/E_2$, only the two of them with smaller $\mathrm{Im} [t']$ will be physically significant. This has been utilised in the orbit classification used for the discussion of holographic interference patterns in section \ref{sec:Coulomb}, which is valid in the regime $E_1/E_2$ large. This will change when $E_1/E_2$ is decreased and the way this changes will depend on $\phi$.

For example, if $\phi=0$ [Fig.~\ref{fig:ret_v_imt}(A)], the field symmetry requires that the steepest descent contour has fixed topology and will not vary by changing $E_1/E_2$ or $\mathbf{p}_0$, in the sense that it prevents the saddles from merging.  There will just be a point that $E_1/E_2$ becomes so small that the contribution of the two (nearly coalescent) saddles with the higher imaginary part may no longer be neglected, however this is a rather ill defined point. Conversely, for $\phi = \pi/4$ [Fig.~\ref{fig:ret_v_imt}(B)], the symmetry around $\mathrm{Re} [t'] = {T}/{4}$ enforces that the steepest descent contour topology must vary with $E_1/E_2$. In fact, for $p_{0\parallel}=0$ there exists a line in the plane ($(E_1/E_2)$ x $p_{0\perp}$) corresponding to the points where two saddles coalesce. At such points the standard saddle point approximation breaks down as one of the assumptions is that saddles are well separated. In its place a uniform asymptotic expansion will be required (see \cite{Faria2002} for an example), which is yet to be developed for the CQSFA. For this reason, the example in this article uses parameters where all important saddle point times are nicely separated. Even for the simple example of (1, 2) fields, there is significant variety in the behaviour of the saddle point times across the parameter space, such that it appears difficult to provide a uniformly applicable orbit classification. However, it is possible to utilise our understanding of saddle symmetries to systematically map out the contour topology, which can help to uncover usable orbit classifications on a case by case basis. This is outside the scope of the present paper and  will be discussed in detail elsewhere.

\begin{figure}[]
\centering
\includegraphics[width=\textwidth]{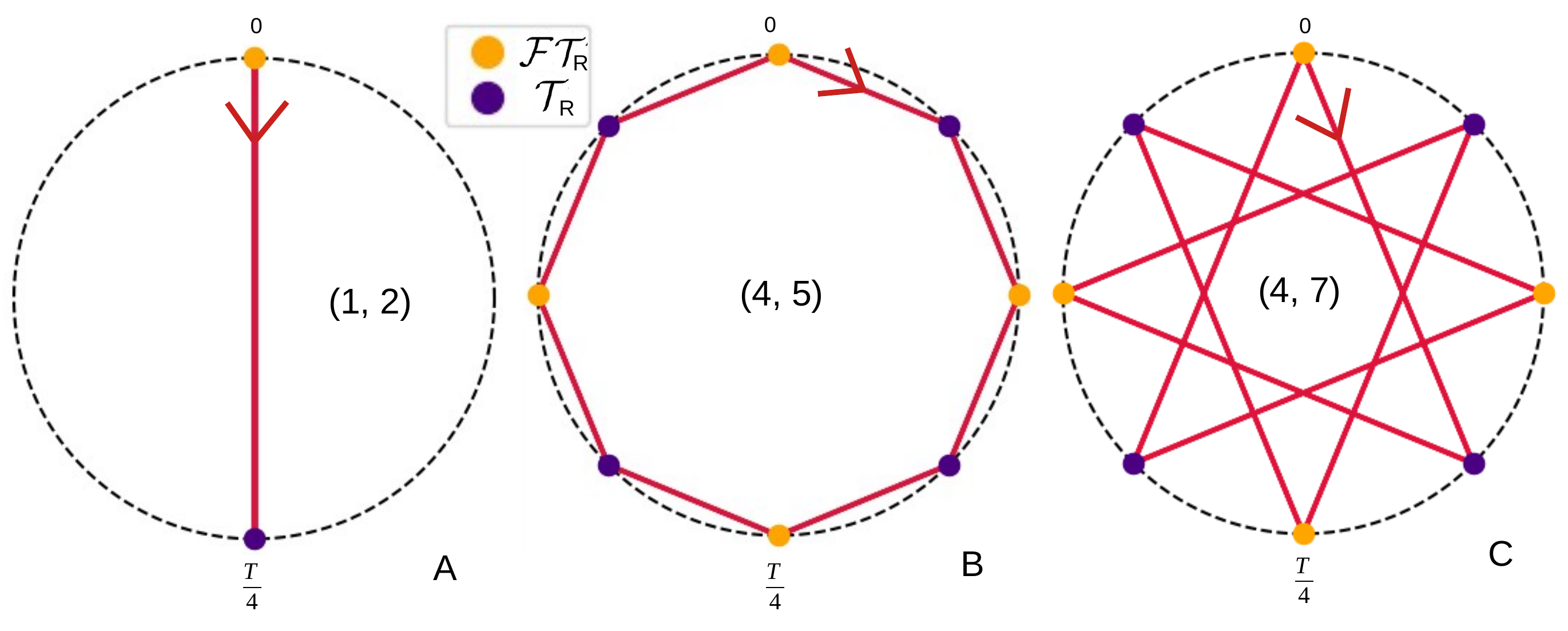}
\caption{\small Schematic representation of phases and axes of symmetry for $(r,s)$ fields, with $r$ and $s$ of opposite parity. The black dashed circles represent the times around which the field could have a symmetry. Each red line segment represents the phase $\phi$ of the field being incremented by ${\pi}/{2s}$ and shows how the symmetry of the field changes under this increment, alternating between $\mathcal{T}_R$ and $\mathcal{F}\mathcal{T}_R$. The yellow (light gray) and purple (dark gray) dots denote $\mathcal{F}\mathcal{T}_R$ and $\mathcal{T}_R$, respectively.}
\label{fig: symm_circles}
\end{figure}
\subsubsection{Phase diagrams}\label{sub:phasediagrams}
In Fig.~\ref{fig: symm_circles}, the phase dependence of equations (\ref{eq: odd even real time solution transformation simple}) and (\ref{eq: odd even imag time solution transformation simple}) is illustrated. The black circle represents a half cycle of the field and the information from equation (\ref{eq: odd even real time solution transformation simple}) translates to the position of the dot on the circle. Concretely, the dot is at the point corresponding to $({nT}/{4})\left[r_s^{-1} + C\left(\frac{1}{r}-1\right)\right]$ where $C$ is given by equation (\ref{eq:C}) and means that for this specific value of $\phi$ there is a symmetry around this time. 
For a field of type $(r, s)$, there are $2r$ possible axis of symmetry within the interval $[0, T/2)$, such that:
\begin{itemize}
  \item Each of these axes become a symmetry axis for some value of the phase $\phi$.
  \item As the phase is incremented by $\pi/(2s)$ the axis of symmetry is shifted by a constant factor, $\frac{T}{4}\left[r_s^{-1} + C\left(\frac{1}{r}-1\right)\right]$,  as given by equation (\ref{eq: odd even real time solution transformation simple}).
\end{itemize}
The previous two points combine to ensure that, as the phase is varied continuously over an interval of width $r\pi/s$, each axis must become a symmetry axis in turn exactly once. This is the phase which shifts the $s$-colored driving wave by a full half cycle. Therefore, if the zero crossings of the $r$ and $s$ driving waves are aligned for $\phi = \phi_0$, they will be aligned again at $\phi = \phi_0 + r\pi/s$ and this will lead to the same axis of symmetry. For a detailed discussion, see Sec.~\ref{sub:fieldsymm} and Fig.~\ref{fig:peak_matching}.
This means that
\begin{eqnarray}
\centering
\frac{T}{4}\left[r_s^{-1} + C\left(\frac{1}{r}-1\right)\right] = \frac{T}{4r}G, \label{eq: generator_cyclic_group} \
\end{eqnarray}
where G must be one of the generators of the cyclic group $\mathbb{Z}/2r\mathbb{Z}$. A similar statement can be made in the case where $r$ and $s$ are both odd
\begin{equation}
\centering
\frac{T}{4}\left[\frac{C}{r} + \frac{1}{2} \right] = \frac{T}{4r}G, \label{eq: generator_cyclic_group_odd} \
\end{equation}
using the factor, $({T}/{4})[{C}/{r} + \frac{1}{2} ]$ from equation (\ref{eq: odd real time solution transformation simple}). The analogous symmetry circles for $(r,s)$ both odd will slightly differ from those in Fig.~\ref{fig: symm_circles} as there will be a contribution from both equation (\ref{eq: odd real time solution transformation simple}) and (\ref{eq: odd real time solution transformation simple 2}), which will lead to two concurrent orbits separated by a quarter of a field cycle. 
The specific symmetry is indicated by the colour of the dot in Fig.~\ref{fig: symm_circles} and this corresponds to information derived from equation (\ref{eq: odd even imag time solution transformation simple}). In particular, if $n$ is even (orange dots), the $\mathcal{F}\mathcal{T}_R$ symmetry holds for the electric field $E(t)$, while for odd $n$ (purple dots), the $\mathcal{T}_R$ symmetry is present. 
For instance, for the (1, 2) field [Fig.~\ref{fig: symm_circles}(A)], $G=1$ so that as we increment $\phi$ by $\pi/4$, the axis of symmetry shifts by $T/4$ each time. In this case the axis of symmetry shifts back and forwards between $t=0$ ($\mathcal{F}\mathcal{T}_R$ symmetry) and $t=T/4$ ($\mathcal{T}_R$ symmetry). For $r=1$, the only possible value of $G$ is 1 since this is the only generator of $\mathbb{Z}/2\mathbb{Z}$. 
However, for $r=4$ the generators of $\mathbb{Z}/8\mathbb{Z}$ are $G=1,~3,~5 \textrm{ and }7$ so the patterns we see in Fig.~\ref{fig: symm_circles} can vary depending on $s$. For $s=5$ [Fig.~\ref{fig: symm_circles}(B)] as the phase is incremented by $\pi/10$ the axis of symmetry shifts by $T/16$.

\begin{figure}[]
\centering
\includegraphics[width=\textwidth]{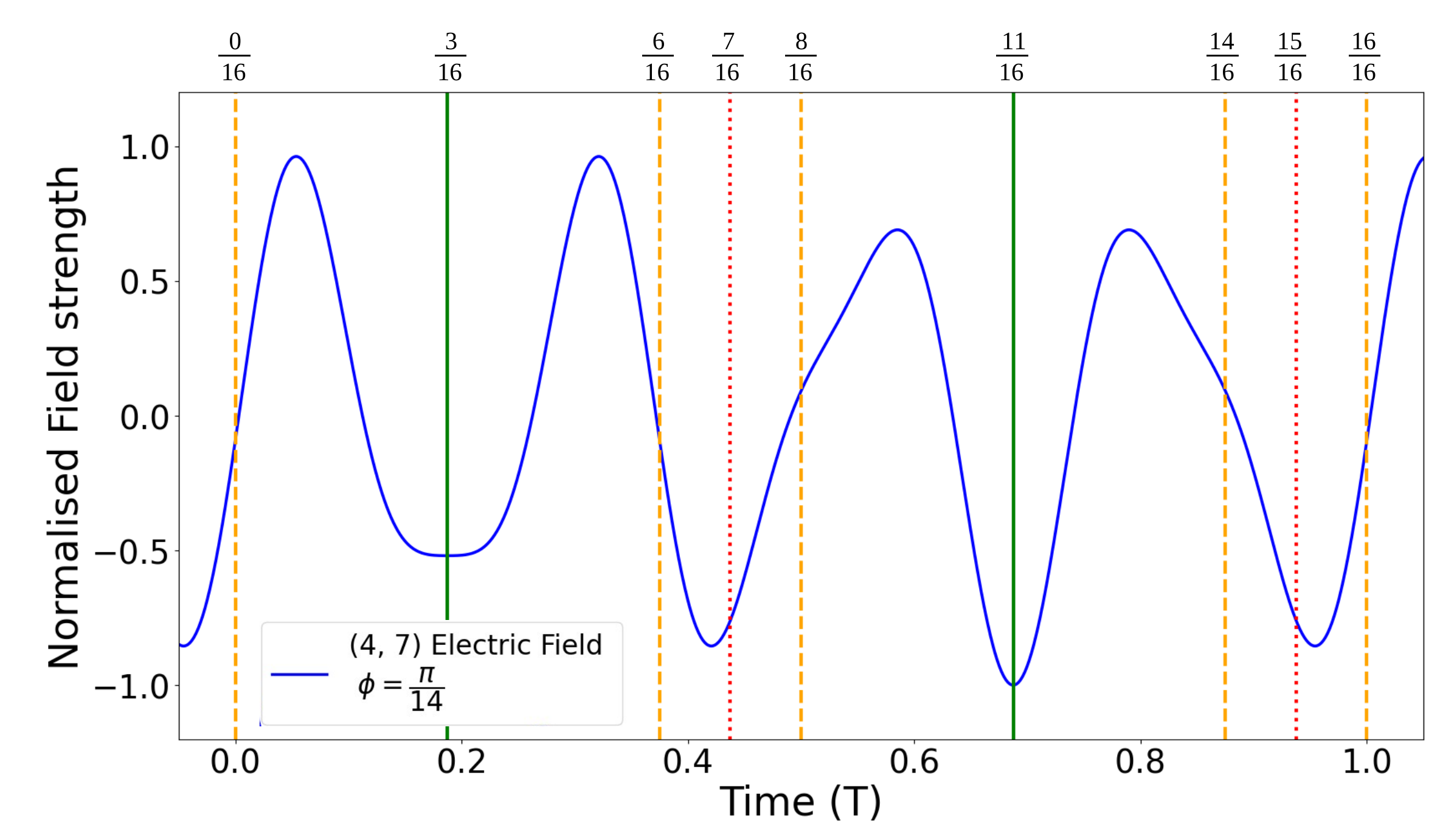}
\caption{\small The temporal profile of the (4,7) field with phase $\phi = \pi/14$ is shown with its symmetry axes marked with green solid vertical lines. This is a specific case of the (4,7) field represented by the symmetry circle in the far right of Fig.~\ref{fig: symm_circles}. The orange vertical dashed lines represent axes which can become symmetry axes of the field under a shift of $\phi$ by $\pi/14$. The red dotted vertical lines represent axes which require the phase to be shifted the maximum amount of $4\pi/14$ in order to become axes of symmetry for the field. This information can be read from Fig.~\ref{fig: symm_circles} by counting the number of blue line segments between different points on the circle. }
\label{fig:symmetry_4_7}
\end{figure}

The situation in Fig.~\ref{fig: symm_circles}(C) can be better understood by looking at the specific $(4,7)$ electric field shown in Fig.~\ref{fig:symmetry_4_7}. Here, $s=7$ means that $G=3$ and as the phase is incremented by ${\pi}/{14}$, the axis of symmetry changes by $3T/16$. This field has $\phi={\pi}/{14}$, so the axis of symmetry can be read from Fig.~\ref{fig: symm_circles} by traversing one blue line segment in the direction of the arrow from zero. This means that the electric field is $\mathcal{T}_R$ symmetric at $t=3T/16$ and in Fig.~\ref{fig:symmetry_4_7} this is represented by a green vertical line. The axes which can be arrived at in Fig.~\ref{fig: symm_circles}  from  a single blue line segment from $3T/16$ are indicated in Fig.~\ref{fig:symmetry_4_7} by orange dashed vertical lines.  These axes are located at $t=0$  and $t=3T/8$. They would be $\mathcal{F}\mathcal{T}_R$ symmetry axes for relative phases $\phi=0$ [$n=0$ in Fig.~\ref{fig: symm_circles}(C)] and $\phi=\pi/7$ [$n=2$ in Fig.~\ref{fig: symm_circles}(C)],  respectively. Similarly, the axes of maximal asymmetry are indicated in Fig.~\ref{fig:symmetry_4_7} by the red dotted vertical lines at $t=7T/16$ and $t=15 T/16$. Starting at $3T/16$ in Fig.~\ref{fig: symm_circles}(C), 4 line segments must be traversed to be reach the axis at $7T/16$. In Fig.~\ref{fig: symm_circles}(C), the final axis of symmetry $t=7 T/16$, where a phase value $\phi=5\pi/14$ means that $\mathcal{T}_R$ will hold is diametrically opposed to those in the starting point.  The same arguments will hold for the second half cycle of the field (see right-hand-side of Fig.~\ref{fig:symmetry_4_7}), although the field shape is different as there is no half cycle symmetry.
On inspection of Fig.~\ref{fig:symmetry_4_7} the field looks ``more" symmetric around the orange axes than it does around the red ones. This is because the procedure discussed above, counting line segments needed to reach a symmetry (a dot in Fig.~\ref{fig: symm_circles}), is an indicator of the value of any reasonably defined asymmetry parameter of the field for this symmetry. A more quantitative explanation can be found in \ref{sec: fieldapp}.

\begin{figure}[]
\centering
\includegraphics[width=100mm]{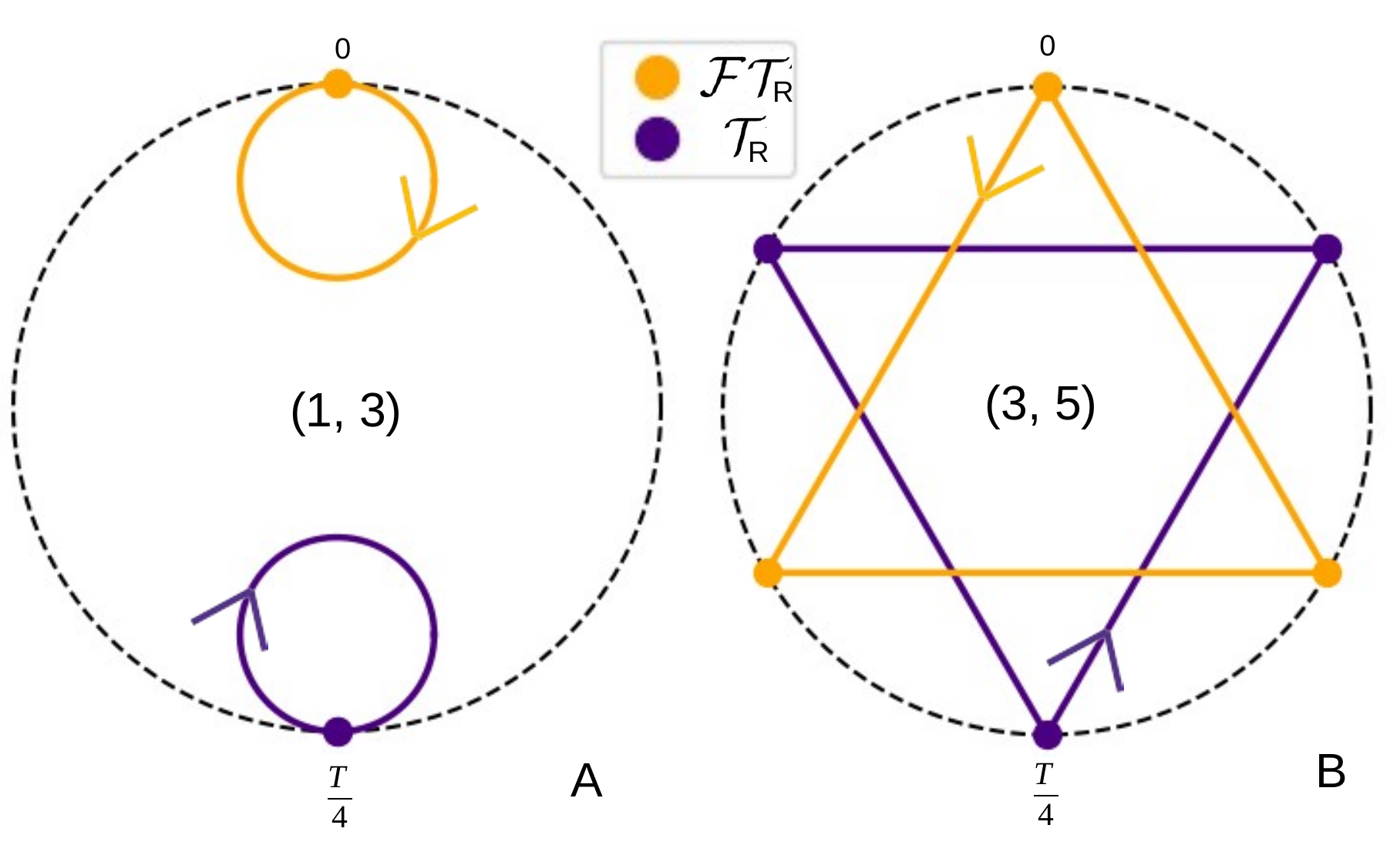}
\caption{\small The black dashed circles and orange and purple dots have the same meaning as in Fig.~\ref{fig: symm_circles}. However, in the case of r and s both odd, the coloured line segments must be interpreted differently. Each line segment (or circle) represents shifting $\phi$ by $\pi/s$. Furthermore, we always have either 0 or 2 symmetries so the two cycles in each panel of Fig.~\ref{fig:sym_circles_odd} must be traversed concurrently. The starting point at $\phi = 0$ is 0 for the yellow cycle and $T/4$ for the purple cycle.}
\label{fig:sym_circles_odd}
\end{figure}

In case both $r$ and $s$ are odd, both symmetries or none exist. However, they occur at different times. For instance, for $\phi=0$, $\mathcal{F}\mathcal{T}_R$ occurs for $t=0 \textrm{ mod } T/2$, while $\mathcal{T}_R$ exists for $t=T/4$ mod $T/2$. Increasing the phase by $\pi/s$ will shift both axes of symmetry simultaneously according to equation (\ref{eq: generator_cyclic_group_odd}). This is exemplified in Fig.~\ref{fig:sym_circles_odd}, in  which the diagram for a (1,3) field is shown in panel A. This is the simplest case and as $\phi$ is increased by $\pi/3$, the field will become symmetric around exactly the same times. The diagram for a (3,5) field is shown in Fig.~\ref{fig:sym_circles_odd}(B). It should be noted that in this case for equation (\ref{eq: generator_cyclic_group_odd}) G=1, however the figure shows cycles which makes 4 steps clockwise for each $\pi/3$ increment of the phase. This is because here the diametrically opposed points represent the same phase but different times, while in Fig.~\ref{fig: symm_circles} this does not hold. Therefore, we can equivalently represent the two cycles as making steps of $G+r$ when we have r and s both odd.

The analysis of the example studied in Section \ref{sec:Coulomb} takes into consideration the existence or non existence of symmetries of $\mathrm{Im}[t']$. This perspective provides an in-depth understanding of how the photoelectron momentum distribution differs for bichromatic fields compared to monochromatic fields in the scenario for which $E_r/E_s$ is large.

\section{Photoelectron momentum distributions}
\label{sec:Coulomb}

\begin{figure}[h!]
\centerline{\includegraphics[width=\textwidth]{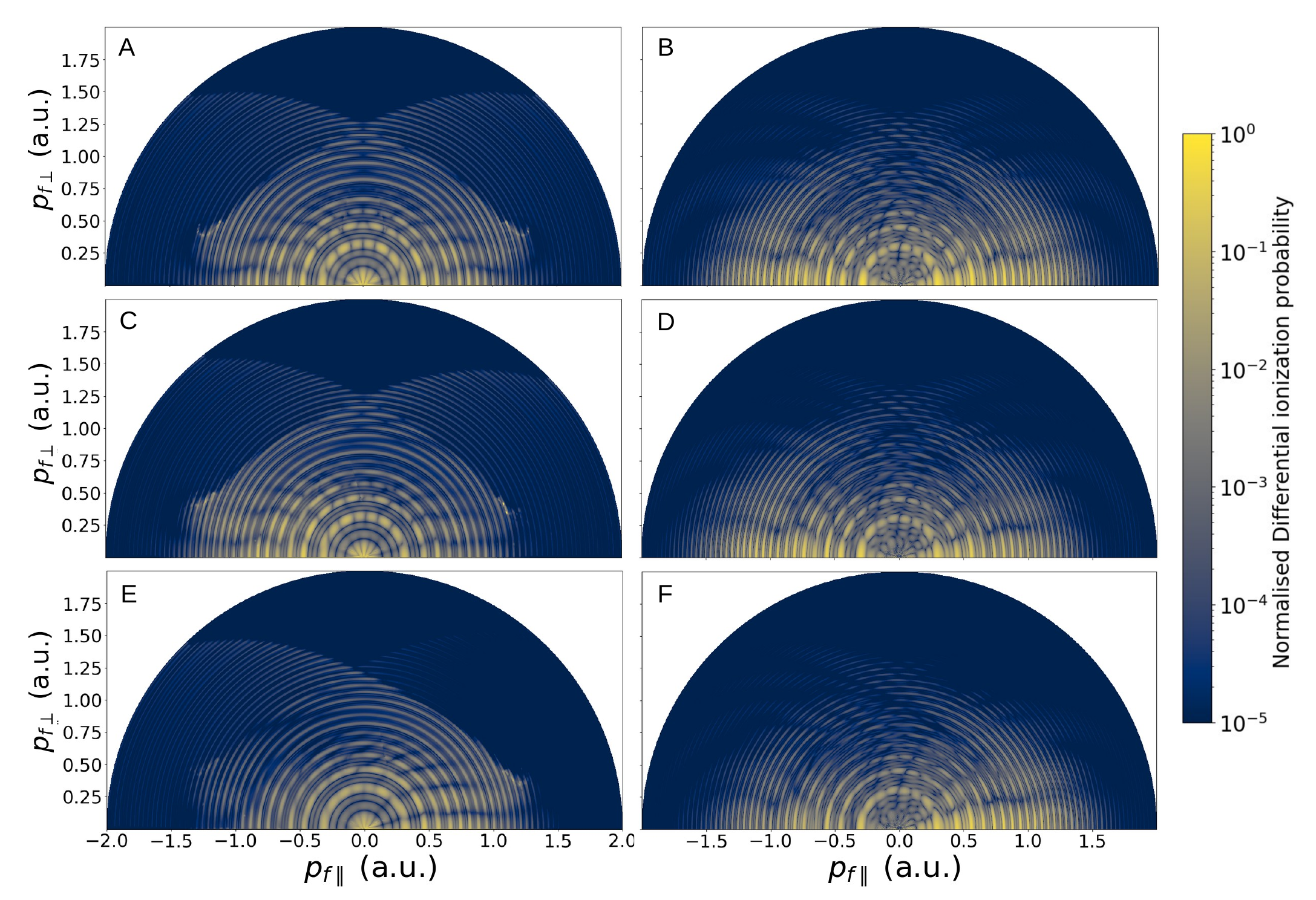}}
\caption{\small Photoelectron momentum distributions calculated for hydrogen ionising from a $1s$ orbital ($I_p=0.5$ a.u.) using the CQSFA and the freely available TDSE solver Qprop\cite{qprop} (left and right columns, respectively). For Qprop, the distributions are created from four cycles of the field with a flat pulse shape and a half cycle on/off period, while for the CQSFA we have considered ionisation events up to four cycles. In order to eliminate residual asymmetries and avoid overpowering ATI rings, in the CQSFA computation we have also performed an incoherent averaging over different unit cells as well as a coherent sum over four cycles (for a discussion see \cite{Werby2021}). In the first row the field is monochromatic, while in the bottom two rows the field is bichromatic of type (1, 2) in equation (\ref{eq:electric_field}). The relative phase differs such that the middle row is for $\phi = 0$ and the bottom row is for $\phi = \pi/4$. The frequency of the monochromatic field and of the fundamental  of the bichromatic field is $\omega=0.057$ a.u. ($\lambda=800$ nm). The intensity of the monochromatic field is 0.0214 a.u. or $1.4~ \times 10^{14}\mathrm{W/cm}^2$ and the bichromatic field intensity is set such that the total ponderomotive energy is the same as that of the monochromatic field. The ratio between the intensities of the two constituent waves is $I_1/I_2=100$, which means that the $2\omega$ wave has an intensity of  $I_{2}=1.4~ \times 10^{12}\mathrm{W/cm}^2$.   The figures are plotted to a logarithmic scale and each panel has been normalized to its maximum value.}
\label{fig:everything}
\end{figure}
In the following, we will provide an example of how the above-stated symmetries and the breaking thereof influence the photoelectron momentum distributions in the presence of the residual Coulomb potential.  We will focus on a bichromatic (1,2) field with relative phases $\phi=0$ and $\phi=\pi/4$, and intensity ratio $I_1/I_2=100$ between the low and high-frequency wave. These are illustrative examples for which the half-cycle symmetry is broken, but which are symmetric with regard to $\mathcal{F}\mathcal{T}_{R}(T/2)$ and $\mathcal{T}_{R}(T/4)$, respectively. The former symmetry will ensure that the field peaks in successive half cycles are equal in magnitude, while the latter will guarantee that the fields gradient around its maximum are of equal magnitude. Due to the presence of the Coulomb potential, the arguments used here are approximate.

We start by comparing the CQSFA results with the outcome of a time-dependent Schr\"odinger equation (TDSE) computation, performed using the freely available software Qprop \cite{qprop,Mosert2016},  for one- and two-color fields [type (1,0) and (1,2) in the present notation]. These results are displayed in  Fig.~\ref{fig:everything} as functions of the final momentum components $p_{f\parallel}, p_{f\perp}$ parallel and perpendicular to the driving-field polarization, and employ the coherent sum of ionization events over four field cycles. 

For the CQSFA, we also perform an incoherent sum over unit cells in order to eliminate residual asymmetries due to artifacts\footnote{In the CQSFA, the ionization times must be restricted to finite ranges. This may lead to artifacts depending how the unit-cell is chosen. In principle, summing over many cycles eliminates this arbitrariness, but also leads to an over-enhancement of ATI rings due to their tending to a Dirac delta comb for monochromatic fields \cite{Becker2002Review}. A way of overcoming such artifacts and retaining the ATI rings without this over-enhancement is to resort to a coherent sum over four cycles of the field and an incoherent average over unit cells similar to that in our previous work \cite{Werby2021}, but modified to incorporate coherent sums over an arbitrary number of cycles. }.  All panels show clear above-threshold ionization (ATI) rings stemming from inter-cycle interference, as well as intra-cycle holographic patterns such as the fan near the ionization threshold, the spider-like fringes near the polarization axis and the interference carpet near the perpendicular momentum axis. The classical ridge associated with rescattering is also present throughout. If a monochromatic field is taken [Figs.~\ref{fig:everything}(A) and (B)], all features are symmetric upon the reflection  $p_{f\parallel} \rightarrow -p_{f\parallel}$ with regard to the perpendicular momentum axis.  For (1,2) fields, this symmetry is lost even for a weak $2\omega$ wave, and the contrast and intensity of the holographic features  is influenced by the relative phase between the two driving waves. For instance, for relative phase $\phi=0$, there is a good contrast in the spider and in the interference carpets, while for $\phi=\pi/4$, the carpets become blurred and the spider loses contrast in the negative momentum region. There are also differences in intensity for the rescattering ridge, which is approximately symmetric for $\phi=0$, but is suppressed for positive momenta if $\phi=\pi/4$ is taken. This suppression is more pronounced for the CQSFA, but is present in all cases. 

\begin{figure}
\centerline{\includegraphics[width=\textwidth]{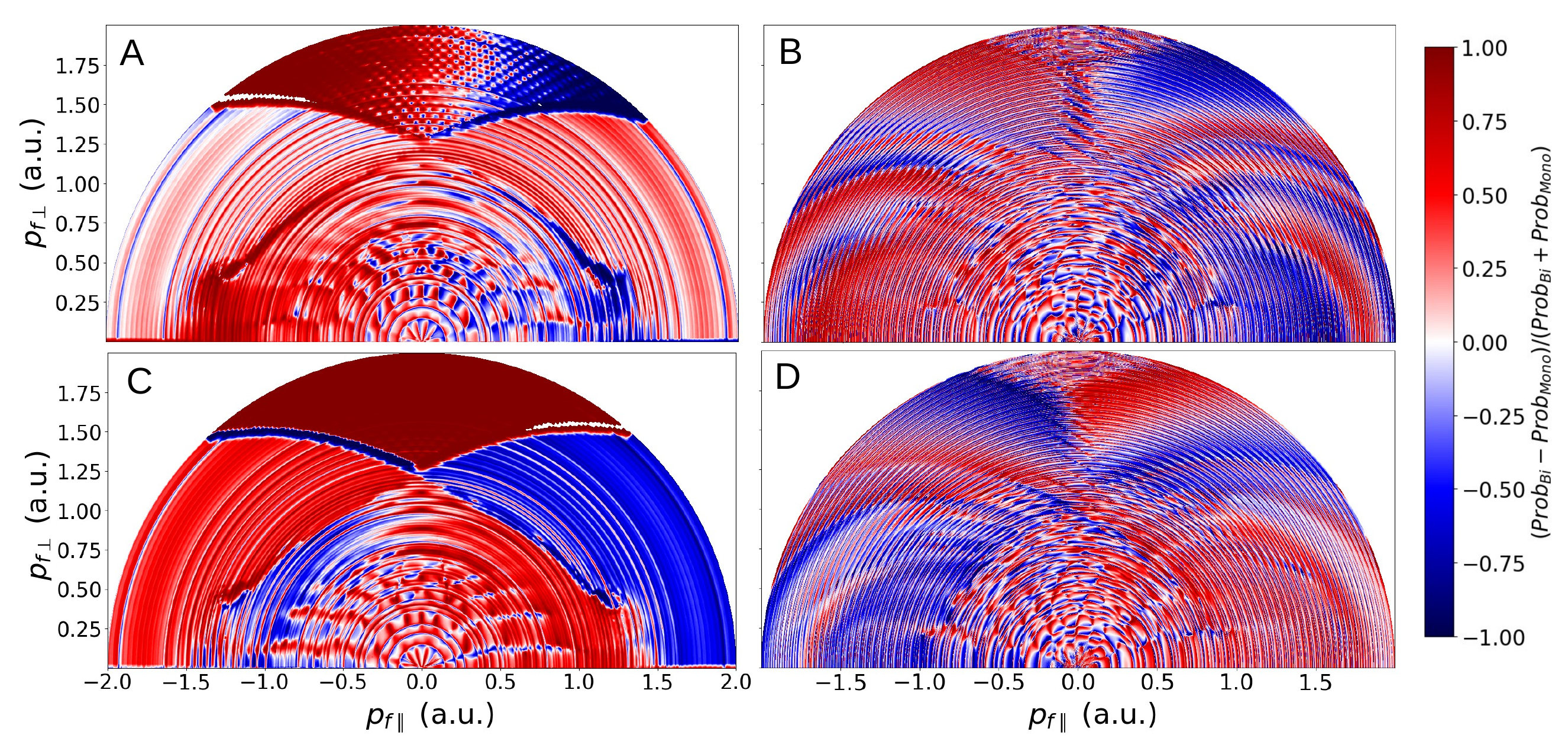}}
\caption{Normalised difference $(|M(\mathbf{p}_f)_{(1,2)}|^2-|M(\mathbf{p}_f)_{(1,0)}|^2)/(|M(\mathbf{p}_f)_{(1,2)}|^2+|M(\mathbf{p}_f)_{(1,0)}|^2) $, where $M(\mathbf{p}_f)$ is given by Eq.~(\ref{eq:CQSFATamp}), between the transition probability for a (1,2) bichromatic field and its monochromatic counterpart, where the plot is scaled by the function $f(x)=\textrm{arcsinh}(\textrm{sign}(x)\sqrt{x})$ and computed for the same field and atomic parameters as in Fig.~\ref{fig:everything}. The upper and the bottom row were calculated for the relative phases $\phi=0$ and $\phi=\pi/4$, respectively, and the left and the right columns correspond to the CQSFA and the TDSE, respectively.  }
\label{fig:differences}
\end{figure}

In order to highlight these asymmetries, in Fig.~\ref{fig:differences} we plot the differences between the bichromatic and the monochromatic field, both for the CQSFA and TDSE computations. Although the results are more pronounced for the CQSFA than for Qprop, overall one can see that the spider, which is symmetric for a monochromatic driving field, is stronger on the left for $\phi=0$ and on the right for $\phi=\pi/4$. Furthermore, the rescattering ridge is stronger for negative (positive) parallel momentum for $\phi=\pi/4$ ($\phi=0$), and the carpet is no longer symmetric for the (1,2) field. For the CQSFA, there are abrupt changes close to the caustic determined by orbit 3, which extends from the perpendicular momentum axis, around $(p_{f\parallel},p_{f\perp})=(0,1.25)$ to roughly  $(p_{f\parallel},p_{f\perp})=(\pm 1.5,0)$, while for Qprop the corresponding shaded areas end almost below the rescattering ridge (see, for instance, blue regions in the negative parallel momentum regions close to the spider legs in Figs. (C) and (D)). These discrepancies and those in Fig.~\ref{fig:everything} are due to additional orbits which coalesce near the rescattering ridge and the caustic around the spider and spiral. Their interference leads to additional structures. These orbits cannot be yet taken into consideration in the CQSFA, as they are likely to require different asymptotic expansions. This is still work in progress and beyond the scope of the present paper. 
 
\begin{figure}[h]
\centerline{\includegraphics[width=\textwidth]{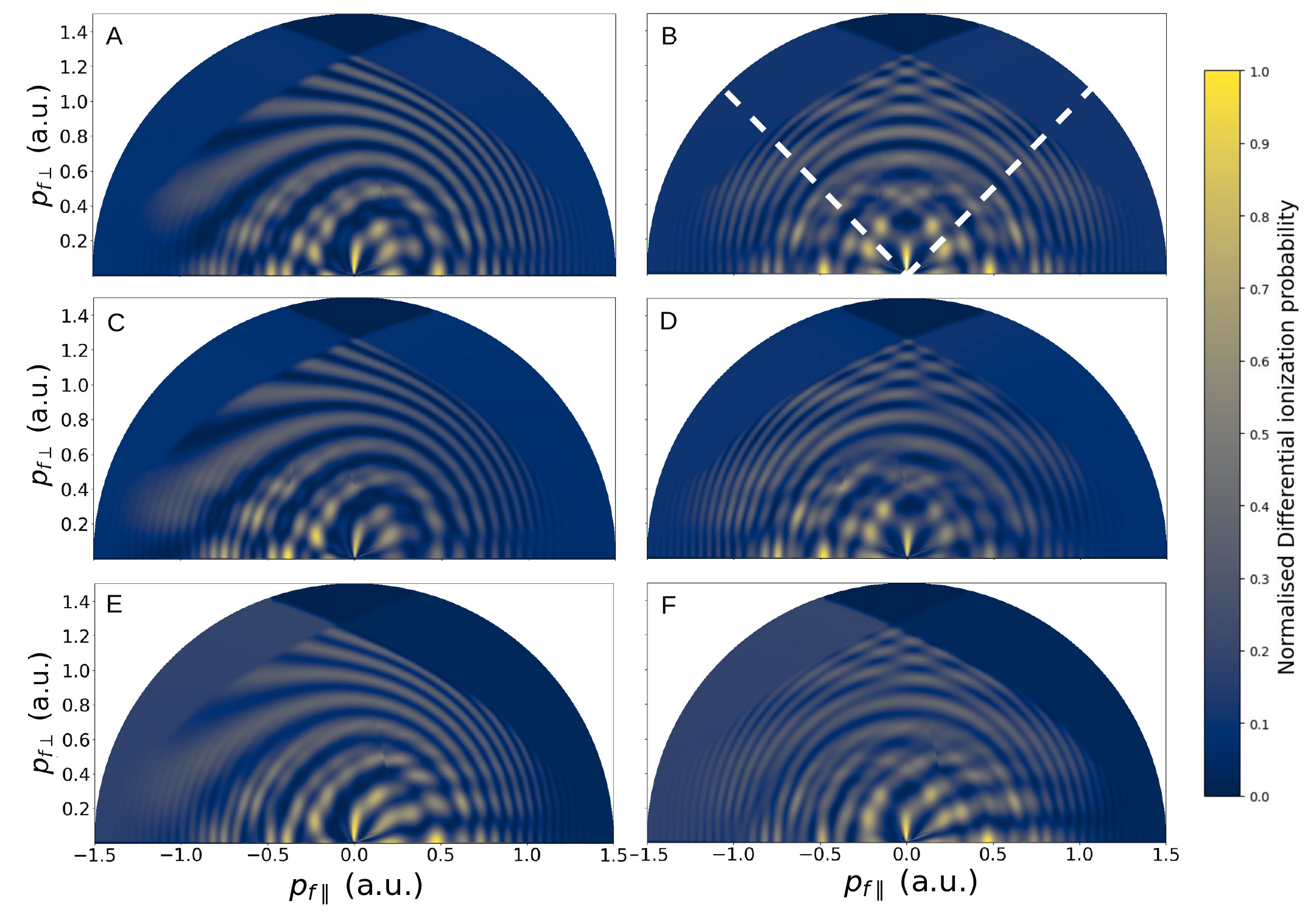}}
\caption{\small Photoelectron momentum distributions with a linear normalised scale calculated for the CQSFA omitting the prefactors and using either a fixed unit cell with $t_{cell}=0$ (left column), or averaging over unit cells according to \cite{Werby2021} (right column). The upper row [panels (A) and (B)] has been calculated for a monochromatic field, and the remaining rows for a bichromatic (1,2) field. In the middle row [panels (C) and (D)] we considered the relative phase $\phi=0$, and in the lower row [panels (E) and (F)] we took $\phi=\pi/4$. The field and potential parameters are the same as in Figs.~\ref{fig:everything} and \ref{fig:differences}. In order to highlight the interference patterns, we display the momentum ranges for which they occur and omit part of the rescattering ridge. The white dashed line in panel (B) indicates the range of momenta used to produce Fig.~\ref{fig: tunnelTimes}.
}
\label{fig:intracycleall}
\end{figure}

Next, in Fig.~\ref{fig:intracycleall}, we have a closer look at intra-cycle interference for fields of different parameters focusing on the CQSFA only. For clarity, 
we omit the prefactors associated with the stability of the orbit and the geometry of the $1s$ state as they lead to additional momentum biases. They improve the agreement with \textit{ab-initio} methods, but sometimes make the effect of quantum interference harder to dissect. We also restrict the ionization events to a single cycle, as the ATI rings can mask some holographic features. The figure shows holographic patterns for monochromatic fields [top row], and bichromatic (1,2) fields with the same relative phases as in the previous figures [middle and bottom rows].  In the left column, we consider a fixed unit cell starting at $t_{cell}=0$, while in the right column we perform an incoherent unit-cell averaging. This eliminates asymmetries due to including a finite range of ionization times, without leading to ATI rings, while fixed unit cells are useful for visualizing specific patterns. The value of $t_{\mathrm{cell}}$ determines the range of time from which saddle points are considered in the coherent sum by introducing an additional phase to the field. Saddle points from the range $[0, 2N\pi/\omega)$ are taken for a field defined by $\mathbf{E}_{r, s, \phi} (t)= [E_0\sin(r\omega t + r\omega t_{\mathrm{cell}}) + E_1\sin(s\omega t - \frac{s}{r}\phi + s\omega t_{\mathrm{cell}})]\mathbf{e}_{||}$ where $N$ is the number of field cycles included. 

Overall, in comparison with the monochromatic field, we see an enhancement of the spiral and spider around the $p_{f\perp}$ axis and in the negative parallel momentum region for relative phase $\phi=0$ [Fig.~\ref{fig:intracycleall}(C)]. For $\phi=\pi/4$ [Fig.~\ref{fig:intracycleall}(E)], there is loss of contrast in the spiral and an enhancement in the spider for $p_{f\parallel}>0$. If unit-cell averaging is performed, one sees that, for monochromatic fields, all patterns are symmetric with regard $p_{f\parallel}=0$ [see Fig.~\ref{fig:intracycleall}(B)]. This includes the interference carpet around the $p_{f\perp}$ axis, which can be seen in the high-energy region of Fig.~\ref{fig:intracycleall}(B), the spider-like fringes close to the $p_{f\parallel}$ axis, and the fan-shaped distributions. This also holds for several features stemming from multi-path interference, such as the structures near the threshold where the fan and the spiral intersect. Once the second field is included, the patterns are no longer symmetric with regard to a reflection around $p_{f\parallel}$, as a consequence of breaking the half-cycle symmetry. Moreover, for $\phi=0$ the carpet and other spiral-related features shift to the left, with an enhancement of the spider in the negative momentum region [Fig.~\ref{fig:intracycleall}(D)], while for $\phi = \pi/4$ these patterns shift to the right and the spider is more prominent for positive parallel momenta [Fig.~\ref{fig:intracycleall}(F)]. Furthermore, the interference carpet has high contrast for $\phi = 0$  but not for $\phi = \pi/4$   [Figs.~\ref{fig:intracycleall}(E) and (F)], while the spider is sharper for $\phi = \pi/4$  and $p_{f\parallel}>0$.  Including prefactors mainly suppresses the spiral and enhances the yield near the polarization axis (see Figs.~\ref{fig:everything} and \ref{fig:differences}).

\begin{figure}[]
\centerline{\includegraphics[width=\textwidth]{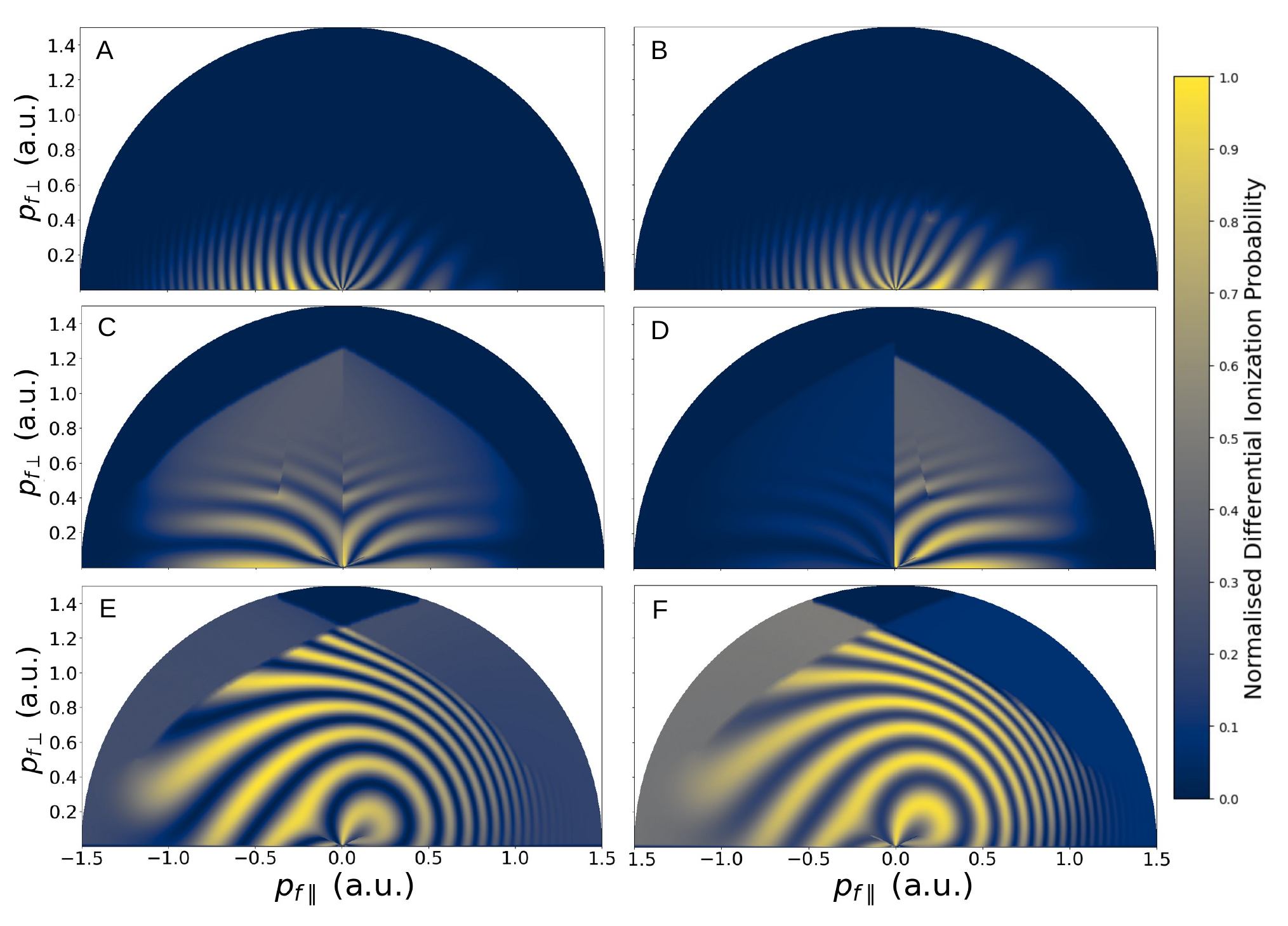}}
\caption{\small Photoelectron momentum distributions computed with the CQSFA for (1,2) fields considering specific pairs of orbits, no prefactor, unit cells starting at $t_{cell}=0$ and the same field and potential parameters are the same as in Fig.~\ref{fig:everything}. In the left column, $\phi=0$ and in the right column $\phi=\pi/4$. For panels (A) and (B), orbits 1 and 2 are coherently summed to give the fan holographic pattern. In panels (C) and (D) the interference of orbits 2 and 3 leads to the spider pattern. Finally in panels (E) and (F) orbits 3 and 4 lead to a spiral pattern. For a more precise notation of how the interfering orbitals are taken into consideration in each momentum region, see Table \ref{tab:Orbits}, where a slightly different classification was introduced to indicate pathways starting at different half cycles. 
}
\label{fig:holpat}
\end{figure}

Due to the half-cycle symmetry being broken for the (1,2) field, we will employ a different orbit classification than in our previous publications \cite{Lai2015a,Maxwell2017}. This classification is stated in Table \ref{tab:orbsbichromatic}, in which the orbits starting in the first half cycle will be indicated by a prime after their number, while for the orbits starting in the second half cycle this prime will be omitted. This classification will be used in the subsequent analysis. In Fig.~\ref{fig:holpat}, we plot the contributions from specific orbit pairs, working under the same assumptions as in Fig.~\ref{fig:intracycleall}. As a further approximation, we consider a specific unit cell starting at $t_{cell}=0$. This sheds more light on how specific patterns are affected.  

\begin{table}
 \centering
 \begin{tabular}{l c c rrrrrrr}
 \hline\hline
 Orbit & $p_{f\parallel}z_0$ & $p_{f\perp}p_{0\perp}$ & $p_{f\parallel}$ 
 \\ [0.5ex]
 \hline\hline
 $1$ &\raisebox{-1.5ex}{+}&  \raisebox{-1.5ex}{+}& +  \\[-1ex]
 $1'$ & & & -- \\[1ex]
 \hline
 $2$ &\raisebox{-1.5ex}{--}&  \raisebox{-1.5ex}{+}& --  \\[-1ex]
 $2'$ & & & + \\[1ex]
 \hline
 $3$ &\raisebox{-1.5ex}{--}&  \raisebox{-1.5ex}{--}& --  \\[-1ex]
 $3'$ & & & + \\[1ex]
 \hline
 $4$ &\raisebox{-1.5ex}{+}&  \raisebox{-1.5ex}{--}& +  \\[-1ex]
 $4'$ & & & -- \\[1ex]
 \hline
 \end{tabular}
 \caption{ Orbit classification for the linearly polarized $(1,2)$ bichromatic field used in this work, generalized from \cite{Yan2010}.  For orbits 1 and 4, the final parallel momentum has the same sign as the tunnel exit, while for orbits 2 and 3 they have opposite signs. Furthermore, for orbits 3 and 4 the transverse momentum component changes sign, while for orbits 1 and 2 they remain unaffected. The last column specifies whether the final momentum is negative or positive, and the prime indicates orbits starting in the first half cycle of the field.  }
 \label{tab:orbsbichromatic}
\end{table}

The upper panels of Fig.~\ref{fig:holpat} display the fan-shaped pattern obtained with the interference of orbits 1 and 2. The intensity of the fan is asymmetric with regard to $p_{f\parallel}=0$, being stronger on the left for $\phi=0$ [Fig.~\ref{fig:holpat}(A)], and on the right for $\phi=\pi/4$ [Fig.~\ref{fig:holpat}(B)]. However, this is a subtle effect, especially for $\phi=\pi/4$. The intermediate panels show far more radical changes for the spider, which arises from the interference of orbits 2 and 3. For $\phi=0$, the spider is slightly weaker for positive parallel momentum, but visible throughout [Fig.~\ref{fig:holpat}(C)], while for $\phi=\pi/4$ it is vanishingly small for negative parallel momentum and very strong for $p_{f\parallel}>0$ [Fig.~\ref{fig:holpat}(D)]. Finally, the main difference in the spiral-like fringes that result from the interference of orbits 3 and 4,  when changing the relative phase from  $\phi=0$ to  $\phi=\pi/4$, is the loss of contrast, which happens throughout but is more pronounced in the negative momentum region. This can be related to the fuzzy interference carpets that occur for $\phi=\pi/4$ in the averaged unit cell case [see Fig.~\ref{fig:intracycleall}(F)].  One should note that the carpet, or the spiral, is due to the interference of orbits that start at different half cycles.  Changing the unit cell or considering an incoherent sum of unit cells just means that the start and end times for ionisation are being shifted. However, the ionisation probabilities and the time difference for events starting at different half cycles are still different. This difference will cause a loss of contrast.  

\begin{table}[h]
\centering
\begin{tabular}{l c c rrrrrrr}
\hline\hline
Structure & Sign of $p_{f\parallel}$ & \multicolumn{2}{c}{Orbits}
\\ [0.5ex]
\hline\hline
 & $+$ &  $2'$ & $3'$  \\[-1ex]
\raisebox{1.5ex}{Spider}& $-$ &  $2$ & $3$ \\[1ex]
\hline
 & $+$ &  $3'$ & $4$ \\[-1ex]
\raisebox{1.5ex}{Spiral} & $-$ 
&  $3$ & $4'$  \\[1ex]
\hline
 & $+$ &  $1$ & $2'$  \\
\raisebox{1.5ex}{Fan} & $-$
&  $1'$ & $2$  \\[1ex]
\hline
\end{tabular}
\caption{Holographic structures (first column), parallel momentum region for which they occur (second column) and the interfering orbits leading to that structure (third column) for the (1,2) fields used in Figs.~\ref{fig:holpat} and \ref{fig: tunnelTimes}. Here, we employ the prime to name the orbits starting in the first half cycle of the (1,2) field. Absence of a prime indicates that the orbits start in the second half cycle. } 
\label{tab:Orbits}
\end{table}

\begin{figure}[h!]
\centering
\includegraphics[width=120mm]{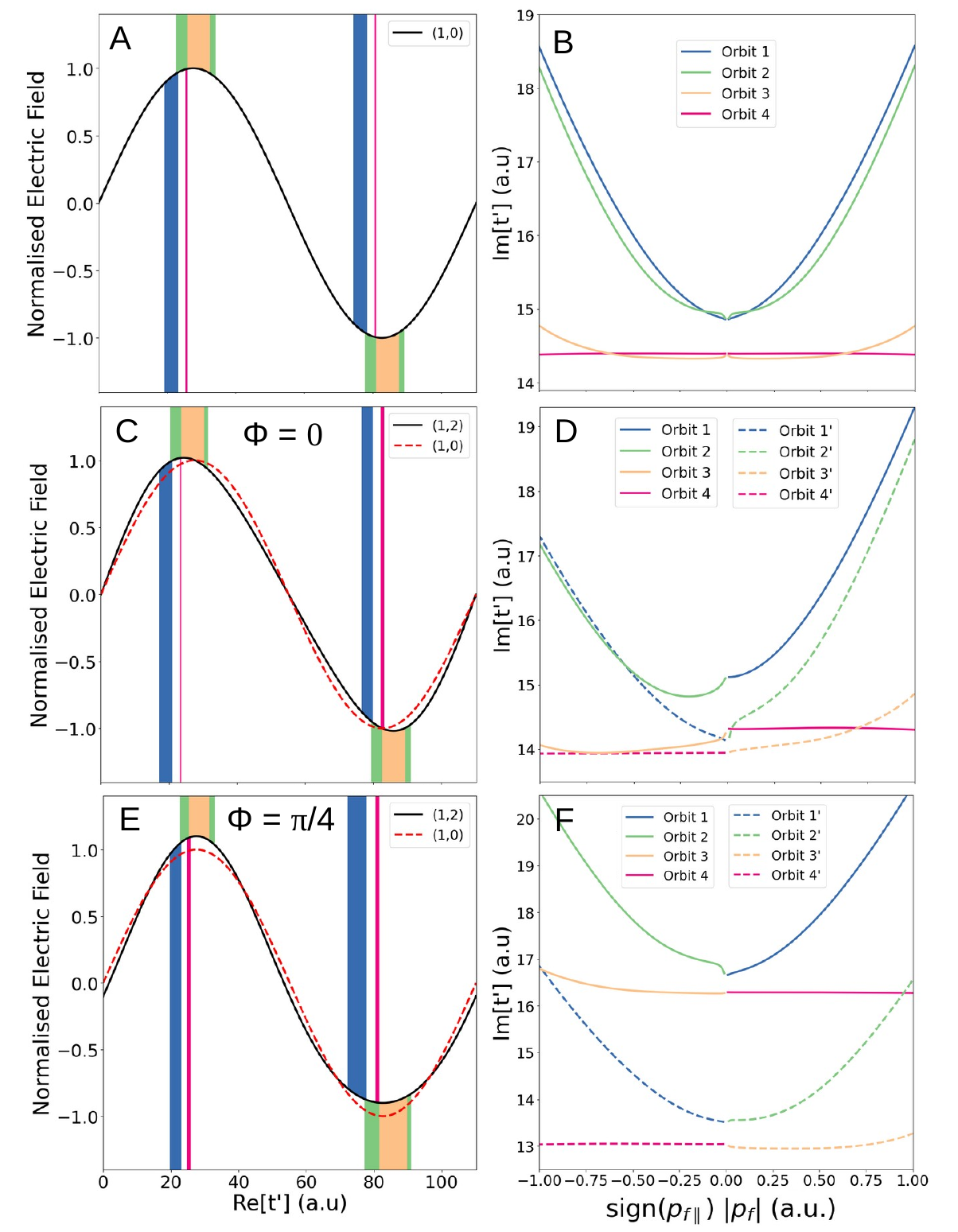}
\caption{\small  The left column shows the temporal profile of the electric field over a single field cycle. For the non-monochromatic fields the monochromatic field is included as a red dashed line for comparison. The range of Re[t'] for each orbit is shown by the width of the coloured blocks. The blue (darkest grey), green (light grey), orange (lightest grey) and pink (dark grey) blocks correspond to orbits 1, 2, 3 and 4 respectively. In the figure, blocks which are above the electric field represent orbits with positive final $p_{f||}$ while those below the electric field have negative final $p_{f||}$. The right column shows the dependence of $Im[t']$ on final $p_{f||}$ for each orbit. For $(1,2)$ fields, the half cycle symmetry is broken. Therefore, in panels D and F dashed and solid lines are used to distinguish between orbits which originate from the first and second half cycles, respectively (see Table \ref{tab:Orbits}). For all panels, the ionisation times considered are those for which the final momentum lies on the radial lines at $\pi/4$ and $3\pi/4$ radians, with $|\mathbf{p}_f|<1$, as represented by white dashed lines in Fig.~\ref{fig:intracycleall}(B). The fields represented in the left column have the same parameters as those used in the Fig.\ref{fig:everything}. The saddle point times in this figure are also determined using the same parameters. 
}
\label{fig: tunnelTimes}
\end{figure}

The specific orbits, as classified in Table.~\ref{tab:orbsbichromatic}, which contribute to each holographic pattern, in different momentum regions are tabulated in Table.~\ref{tab:Orbits}. As in Table .~\ref{tab:orbsbichromatic}, the prime indicates the orbits which start in the first half cycle. In order to clarify the behaviour of these interfering orbits, in Fig.~\ref{fig: tunnelTimes}, we plot the temporal profiles of the electric field, together with the imaginary parts of the ionization times (left and right columns, respectively) as functions of the final parallel momentum keeping the perpendicular momentum component fixed. 

These times can be associated with the dominant term in $\mathrm{Im}[S]$, which dictates the ionization probability associated with a specific electron pathway. Because the ionization probability scales with $\exp[-2\mathrm{Im}[S]]$, and the dominant term in the action  is proportional to $it'$, the smaller $\mathrm{Im}[t']$ is, the more probable tunneling will be. This will lead to a particular quantum pathway being prominent. For examples see \cite{Maxwell2017,Maxwell2017a} in the context of photoelectron holography, as well as our previous work on laser-induced nonsequential double ionization \cite{Liu2004,Faria2004b,Faria2009,Faria2012} and molecular high-order harmonic generation \cite{Faria2007}. For interpreting the following picture, it is useful to note that an electron leaving at the peak of the field is expected to have vanishing momenta. In the SFA, $p_{f\parallel}=0$ also gives the most probable momentum for ionization to occur. However, due to the presence of the Coulomb potential the conservation of momentum is lost. This means, for instance, that an electron leaving with vanishing momentum along orbit 1 would be decelerated and trapped by the potential, while an electron leaving along orbit 2 or 3 would be accelerated by it. A more complete analysis of the final to initial momentum mapping is provided in our early work \cite{Lai2015a,Maxwell2017,Maxwell2017a} (see also \cite{Li2014} for a forward momentum mapping, albeit with a different orbit classification). Here, however, we are interested in the final momenta as they will determine which trajectories contribute to the interference patterns. 

For comparison, in the upper row of Fig.~\ref{fig: tunnelTimes} the monochromatic-field scenario is displayed. The most pronounced feature is that, for a monochromatic field, the curves are symmetric around the $p_{f\perp}$ axis. This is a consequence of the half-cycle symmetry of the field: the reflection $\mathbf{p}_f \rightarrow -\mathbf{p}_f$ corresponds to events displaced by half a cycle, for which, apart from a minus sign, the field and its gradient are identical [see Fig.~\ref{fig: tunnelTimes}(A)]. 
Still, the curves in Fig.~\ref{fig: tunnelTimes}(B) behave in distinct ways: For orbit 1,  $\mathrm{Im}[S]$ exhibits a single minimum at $p_{f\parallel}=0$, orbits 2 and 3 exhibit minima at non-vanishing momenta and orbit 4 has a much flatter behaviour.
These features have been discussed elsewhere \cite{Maxwell2017} and are due to the orbits' dynamics. Orbit 1 reaches the detector directly, and is decelerated by the binding potential, thus behaving similarly to its SFA counterpart. Orbits 2 and 3 are field-dressed hyperbolae, so that they are accelerated by the Coulomb potential when they are near the core. This renders the most probable final momentum non-vanishing. Finally, for orbit 4 $\mathrm{Im}[t']$ is very small and much flatter, with regard to the electron momenta, than those for the other orbits. This happens because the ionization times associated with this orbit are located within a very narrow range around the maxima of the field. Therefore, the ionization probability associated with it will be high and vary much less with the electron momentum. For details see our recent manuscript  \cite{Werby2022}.

The half-cycle symmetry is broken when the $2\omega$ field is added (remaining panels), so that the imaginary parts $\mathrm{Im}[t']$ associated with the orbits starting at different half cycles are no longer degenerate. Nonetheless, this behavior is different for $\phi=0$ and $\phi=\pi/4$. For $\phi=0$ [Fig.~\ref{fig: tunnelTimes}(D)], this degeneracy is broken in two main ways. First, there are vertical shifts, due to the effective potential barriers being different, and a `tilting' around $p_{f\parallel}=0$ associated to the field gradients being different around a field extremum, with trajectories starting at different half a cycle leading to strikingly different slopes in for $\mathrm{Im}[t']$. These features can be spotted very clearly around $p_{f\parallel}=0$, and can be associated with the fields and corresponding ranges for $\mathrm{Re}[t']$, displayed in Fig.~\ref{fig: tunnelTimes}(C)].  The figure shows that, due to the reflection symmetry around the field zero crossing ($\mathrm{Re}[t']=T/2$), the field extrema remain the same up to a minus sign, while the field gradients around each field maxima or minima differ. This will lead to the different slopes in the imaginary parts of $t'$, but not so pronounced vertical shifts.
In contrast, for $\phi=\pi/4$, the vertical shifts in $\mathrm{Im}[t']$ are much larger [Fig.~\ref{fig: tunnelTimes}(F)]. This is caused by the field extrema at consecutive half cycles having different amplitudes [see Fig.~\ref{fig: tunnelTimes}(E)]. However,  the curves $\mathrm{Im}[t']$ associated with processes starting at different half cycles look very similar, apart from a reflection around $p_{f\parallel}=0$. This is due to the reflection symmetry around $T/4$, which leads the field having the same gradient, in absolute value, around maxima or minima. 
The smaller displacement in $\mathrm{Im}[t']$ for orbits starting at consecutive half cycles explains why the differences in intensity observed for the holographic patterns are subtler in the $\phi=0$ case. 

Next, we analyse the behavior of specific holographic structures using Fig.~\ref{fig: tunnelTimes}. 
Fig.~\ref{fig: tunnelTimes}(F) suggests that, for $\phi=\pi/4$, the contributions starting in the second half cycle will be strongly suppressed due to $\mathrm{Im}[t']$ being large. This will lead to an extremely weak spider for $p_{f\parallel}<0$, as it is associated with orbits 2 and 3 in Fig.~\ref{fig: tunnelTimes}(F). It will also cause a loss of contrast in the fan and the spiral, as those patterns result from the interference of pathways starting at different half cycles. For instance, in the $p_{f\parallel}<0$ region, the contributions of orbit $2$ and $3$ are very weakened. This means that, when orbit $2$ interferes with $1'$ to form the fan, or when orbit $3$ interferes with $4'$ to form the spiral, the interference fringes will be blurred. A similar argument can be used for the weakened orbit $1$ interfering with $2'$ and for the suppressed orbit 4 interfering with $3'$, in the $p_{f\parallel}>0$ region. The suppression of the rescattering ridge for the positive parallel momentum region is also associated with $\mathrm{Im}[t']$ being large for orbit $4$.
For $\phi=0$, Fig.~\ref{fig: tunnelTimes}(D) shows that the asymmetries are subtler as they are caused by smaller shifts and by the field gradients around its extrema being different. For instance, the spiral and the spider are sharp and comparable throughout, with the spider (spiral) slightly stronger on the left (right). A noteworthy feature is the suppression of the fan's contrast and strength for $p_{f\parallel}>0$, due to the steeper gradients of orbits $1$ and $2'$ and larger differences in $\mathrm{Im}[t']$ in this region. Flatter, almost merging $\mathrm{Im}[t']$ for $1'$ and $2$ leads to a sharper contrast and a stronger fan for $p_{f\parallel}<0$.

Due to the presence of the Coulomb potential, the above discussion only holds approximately. The interplay between the driving field and the Coulomb potential is highly non-trivial, and even small changes in the electron's binding energy can lead to qualitatively different behaviors for the CQSFA (for a recent example see \cite{Bray2021}).

\section{Conclusions}
\label{sec:conclusions}

In the present work, we perform a systematic analysis of the symmetries present in strong-field ionization with linearly polarized $r\omega-s\omega$ bichromatic fields of commensurate frequencies, and for what field frequency ratios and relative phases they are broken or retained. Apart from the well-known half-cycle symmetry, which is broken if $r+s$ is odd, there are temporal reflection symmetries around the field maxima and zero crossings. These three symmetries are always present for linearly polarized monochromatic fields. However, for $(r,s)$ fields this is not always the case. For instance, if $r+s$ is even, the half-cycle symmetry will not be broken, but, depending in the dephasing between both waves, the above-mentioned reflections may or may not hold. On the other hand, if the half-cycle symmetry is broken as in the case where $r+s$ is odd, at least one of the other two is broken as well. We provide explicit values for the relative phase $\phi$ for which symmetries exist, for a general linearly polarized $(r,s)$ field, together with the corresponding times for which they occur. This is performed in a saddle-point framework, and entails not only the real parts of the ionization times, but also their imaginary parts. This will have direct consequences in the contrast and prominence of holographic patterns.

We analyze these effects in the theoretical framework of the Coulomb quantum-orbit strong-field approximation (CQSFA), using saddle-point methods and investigating how the symmetries influence the existing ionization times and subsequent orbits. Saddle-point equations provide valuable insight not only on the real parts of the ionization times, which can be associated with electron orbits and their interference, but also on its imaginary parts, which are related to ionization probabilities and therefore clarify how prominent a specific process is.  Furthermore, we provide an example of how symmetry breaking influence specific holographic patterns, such as the fan, the spider and the spiral, for  $\omega-2\omega$ fields. Throughout, we considered a weak high-frequency wave, so that the propagation in the continuum is minimally disrupted.  Together with breaking the half-cycle symmetry, breaking at least one of the reflection symmetries mentioned above causes changes in contrast and/or strength for specific holographic patterns. If the reflection $\mathcal{F}\mathcal{T}_R(T/2)$ with regard to field zero crossing is unbroken, the field gradients will be different but the peaks will not differ in absolute value. This means that patterns starting in different half cycles will mostly retain their sharpness, but the spider will lose contrast, as it stems from orbits starting in the same half cycle. On the other hand, breaking the reflection $\mathcal{T}_R(T/4)$ with regard to the field peaks, will lead to loss of contrast in the fan and spiral, but the spider will remain sharp. However, it will lose intensity for the half cycle in which the field is weaker.  These results also show that the orbit classification introduced in \cite{Yan2010} is not sufficient to deal with linearly polarized fields that are not monochromatic or cannot be approximated by a monochromatic wave, and will depend on the field frequencies, relative intensities and relative phase. In the specific example provided here, eight types of orbits were necessary to interpret the CQSFA outcome. 

A previous publication \cite{Porat2018} has also reported alterations in the contrast of holographic patterns in two-color $(1,2)$ fields, which resulted from critical changes in tunnel ionization probabilities. These features were observed experimentally for argon atoms using a fundamental laser ﬁeld with a
wavelength of $\lambda = 788$ nm and an intensity $I= 1.3 \times 10^{14}~\mathrm{W/ cm}^{2}$. However, therein the main emphasis was on breaking the half-cycle symmetry, its influence on the spider and how this can be used for inferring ionization times. Furthermore, the standard SFA was used and the $2\omega$ field was  treated as a perturbation. Interestingly, similar effects to those in our work are reported, which suggest that the influence in ionization comes mainly from the laser field and its properties. Nonetheless, we show that the residual Coulomb potential plays an important role in determining the relevant sets of orbits and holographic patterns. We anticipate that, for bichromatic fields with driving waves of comparable intensities, the Coulomb potential will become even more important due to the presence of secondary, less prominent field maxima. They are expected to lead to other types of relevant orbits, which will be more critically affected by the potential. However, a detailed study of such features is beyond the scope of the present work. A key issue is to deal with the changes in the saddle-point contours and number of relevant orbits that occur in this parameter range, for a Coulomb-distorted approach such as the CQSFA. This may require the development of novel asymptotic approaches, apart from the uniform approximation for two nearly coalescent saddles that is widespread in strong-field physics \cite{Faria2002}. Therefore, it is not clear whether the number, relevance and types of orbits can be determined for a \textit{general} linearly polarized $(r,s)$ field of arbitrary frequency and intensity ratios, and relative phase. For experimental studies of photoelectron holography in two colour fields of comparable intensity see \cite{Xie2012, Arbó2015}.

Other important questions, which will be the topic of further investigations, are whether one may use tailored fields to manipulate holographic structures and thus extract information about the target which will not be available otherwise. A key difficulty is that some holographic patterns obfuscate others, so that the features of interest may be difficult to extract. For instance, the spider is quite prominent, such that, in early experiments, it was necessary to subtract its influence in order to see a subtler, fishbone structure \cite{Bian2012,Haertelt2016,Li2015}. The spider also obscures a spiral-like structure except close to the perpendicular momentum axis. The spiral arises from orbits that strongly interact with the core, and therefore is a promising holographic tool \cite{Maxwell2020,Peng2021}. Orthogonally polarized fields are potentially powerful tools for disentangling different types of intra-cycle interference, but their influence has been mostly interpreted using the Coulomb-free, standard SFA \cite{Li2016,Xie2017}. The same holds for the understanding of dynamical symmetries of the field and the target: although there is a considerable body of work in this direction \cites{Habibovich2021}, the influence of the residual potential remains largely unexplored. 

A key challenge is that the presence of residual potentials leads to  deviations from the simple mapping $\mathbf{p}_0=-\mathbf{A}(t')$ dictated by the SFA. This has been pointed out for orthogonal two color fields \cite{Zhang2014}, but is expected to happen as soon as the acceleration caused by the potential in the continuum becomes significant. Therefore, many of the arguments employed in the present work are approximate. If one is dealing with highly directional states or non-isotropic potentials, this becomes a non-trivial matter. In the SFA framework, aligning an axis of symmetry of the system with the field will guarantee that this symmetry is retained \cite{Habibovich2021}, while in the presence of the residual potential this will not necessarily hold.  It is not yet clear how it will affect targets with specific geometries and/or internal degrees of freedom, which, \textit{per se}, cause phase changes and modulations in holographic structures \cite{Fernandez2009,Meckel2014,Liu2016A,He2018,Walt2017}. 

For instance, recently, we have found that the presence of a central potential may move the electron dynamics away from the polarization axis for orbit 4 \cite{Bray2021}, which led to a prominent rescattering ridge for excited Helium even if the initial bound state  $2p$ was oriented perpendicular to the driving-field polarization. This effect  was extremely sensitive to the binding energy and the tunnel exit, and went against what one would expect from the physical intuition provided by the SFA. Furthermore, depending on the parity of the initial orbitals, some holographic structures will pick up extra phase differences. In order to assess that, one must minimize the continuum phase differences and Coulomb distortions by using a molecule and a companion atom with very similar ionization potentials \cite{Kang2020}.

If this is unclear for non-isotropic excited states in atoms, for molecules we expect this to be even more extreme. A molecular potential and also molecular orbitals are in general highly directional, with an angular dependency, symmetry axes, and other issues that critically affect holographic patterns.  For instance, in \cite{Meckel2014} it was shown that the holographic structures are very sensitive with regard to the molecular orientation, which can be used to generate phase offsets. Signatures of different bound states \cite{He2018} and nuclear-electronic coupling \cite{Walt2017} also have a strong influence on photoelectron momentum distributions.  A forward-backward asymmetry along the polarization vector for photoelectron spectra in $H_2^+$ can also be caused by the population of degenerate continuum states with opposite parities \cite{Fernandez2009}.

Specifically with regard to molecular systems and trajectory-based models, there may be intramolecular trajectories, which may move from one center to the other without reaching the continuum. Early high-order harmonic studies have shown that, even within the SFA these orbits may lead to quite prominent features \cite{Kopold1998}. In the presence of the binding potential, these orbits are expected to be chaotic or lead to resonances, population trapping and threshold effects.  Other types of orbits may leave from one center and rescatter off the other \cite{Chirila2006,Faria2007}. This may require a  different orbit classification for the CQSFA orbits than that employed here. 
 Finally, one should bear that the issues mentioned above are based on a highly simplified, one-electron picture. In reality, however, the core dynamics and multielectron effects must also be taken into consideration \cite{Smirnova2009}.
The present work is intended as a step towards the understanding of symmetry in a Coulomb-distorted context, an in providing a theoretical framework for a generic $(r,s)$ two-color linearly polarized field.

\textbf{Acknowledgements: }We would like to thank A. C. Bray and A. S. Maxwell for useful discussions. This work was partly funded by grant No.\ EP/J019143/1, from the UK Engineering and Physical Sciences Research Council (EPSRC). The authors acknowledge the use of the freely available QPROP software (www.qprop.de).

\appendix
\section{Quantifying field asymmetries}
\label{sec: fieldapp}
In the specific case of a two-colour field, one may define asymmetry parameters in terms of how greatly the phase $\phi$ differs from the value of $\phi_{\textrm{s}}$ when the symmetry holds, or by using the corresponding electric field. 

\begin{figure}[h!]
\centering
\includegraphics[width=\textwidth]{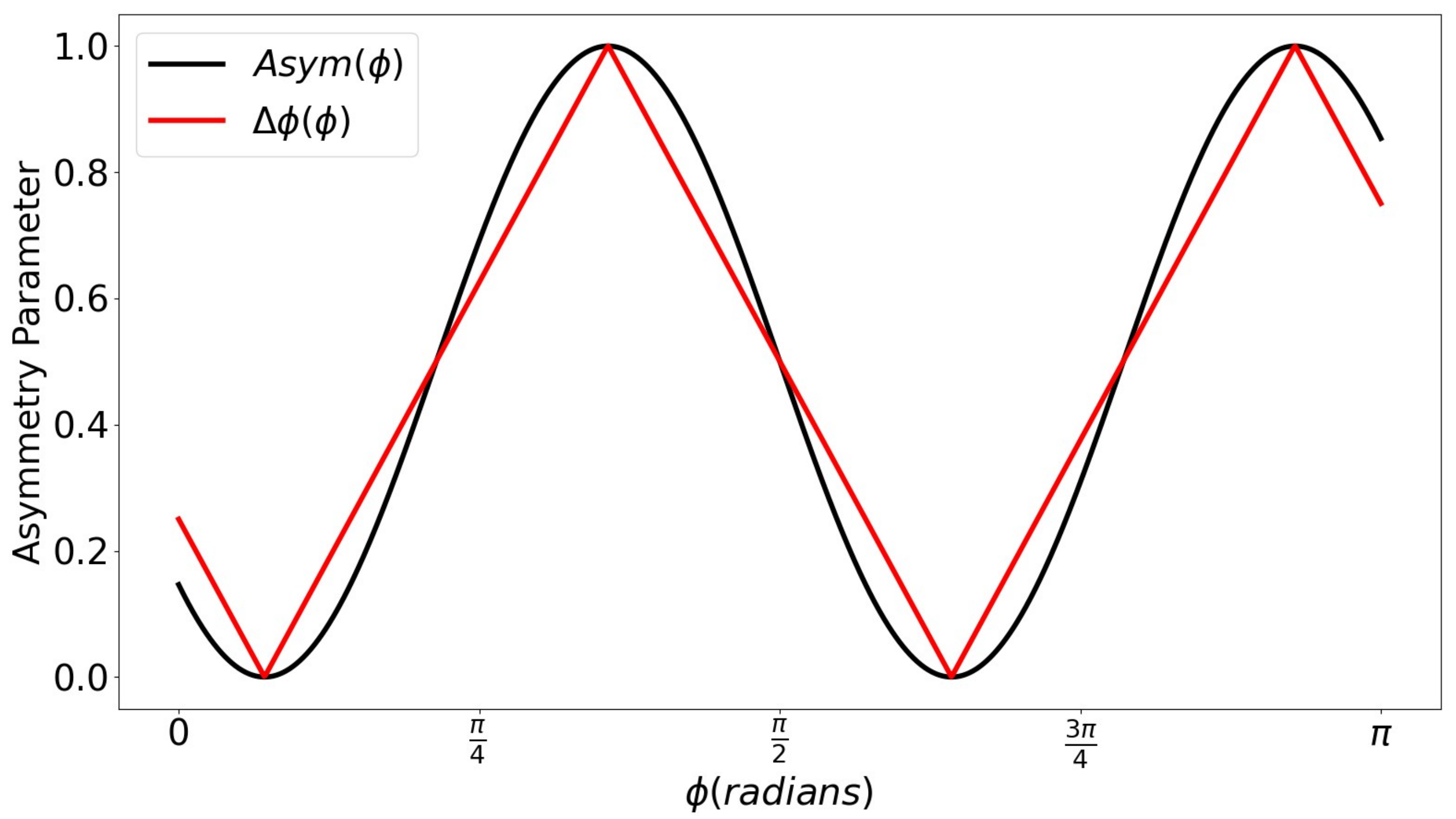}
\caption{\small Two normalised asymmetry parameters are shown for the (4, 7) field. The specific symmetry being measured is that associated with the reflection $\mathcal{T}_R(3T/16)$ which is the symmetry that holds for the field in Fig.~\ref{fig:symmetry_4_7}. The asymmetry parameters are equivalent in that they preserve inequality.  
}
\label{fig:asymmetry}
\end{figure}

An example of a reasonable asymmetry parameter for the transformation $\mathcal{T}_R(\tau)$ is 
\begin{equation}
\centering
\mathrm{Asym}(\phi) = \int^{2\pi}_0 (E_{r,s,\phi}(t) - \mathcal{T}_R(\tau)E_{r,s,\phi}(t))^2 dt. \label{eq:asymmetry_comparison} \
\end{equation}
This parameter vanishes if the field is symmetric upon $\mathcal{T}_R(\tau)$, and  is equivalent to a parameter defined just from the phase $\phi$,
\begin{equation}
\centering
\Delta\phi(\phi) = \textrm{min}_{\phi_{\textrm{s}}}(|\phi-\phi_{\textrm{s}}|) \label{eq:delta_phi} \
\end{equation}
where $\phi_{\textrm{s}}$ is the set of $\phi$ such that a specific symmetry holds, in the sense that inequalities are preserved when the parameters are interchanged [Fig.~\ref{fig:asymmetry}]. This means that for all $\phi_1$, $\phi_2$,
\begin{equation}
\centering 
\textrm{Asym}(\phi_1)<\textrm{Asym}(\phi_2) \iff \Delta\phi(\phi_1) < \Delta\phi(\phi_2).
\label{eq:inequality} \
\end{equation}

This must hold by considering the conditions for symmetries to exist in Fig.~\ref{fig:peak_matching} and the fact that sinusoidal functions vary monotonically between their zeroes and the midpoint of that zero and the adjacent zero. We illustrate both asymmetry parameters in Fig.~\ref{fig:asymmetry}, taking into consideration $\mathcal{T}_R(3T/16)$ and a $(4,7)$ field. The figure shows that these parameters vanish for relative phase $\phi=\pi/14$ and $\phi=9\pi/14$, which is consistent with the symmetry upon the reflection $\mathcal{T}_R(3T/16)$ holding for these phases. The parameter also reaches its maximum for $\phi=5\pi/14$, which is consistent with the discussion of Fig.~\ref{fig:symmetry_4_7}.

\section{Determining saddle symmetries}\label{sec: sadapp}
The best way to discuss symmetries of the saddle-point solutions is to take some solution of equation (\ref{eq: general two colour spe}), $x_\nu$, and find some transformation of this solution such that the result of such transformations, $x_\mu$, is equally a solution of equation (\ref{eq: general two colour spe}).

First consider just the case where the pair $(r, s)$ have opposite parity so we consider phases $\phi = {n\pi}/{2s}$ such that $n \in \mathbb{Z}$. Take the complex conjugate of equation (\ref{eq: general two colour spe}) to give
\begin{equation}
\centering
\eqalign{\frac{E_s}{s}\left[\frac{\tilde{x}_\nu^s}{e^{-i\phi s/r}}+\frac{e^{-i\phi s/r}}{\tilde{x_\nu}^s}\right] + \frac{E_r}{r}\left[\tilde{x_\nu}^r+\frac{1}{\tilde{x_\nu}^r}\right] \\ - 2\omega\left[p_{0\parallel} \pm i\sqrt{2I_p+p_{0\perp}^2}\right] = 0,}\label{eq: odd even solution step 1 appendix} \
\end{equation}
where the tilde denotes complex conjugation. By introducing the number $d$, we can construct $e^{i\pi n(d-s)/r}e^{-i\pi n(d-1)/r}e^{i\pi n(s-1)/r} = 1$. For $d$ such that $({d-1})/{s} = q \in \mathbb{Z}$, $e^{i\pi n}$ and $e^{i\pi n(d-1)/s}$ are equal to either plus or minus one. Likewise, if $({d-s})/{r} = k \in \mathbb{Z}$, $e^{i\pi n(d-1)/s}$ will be either plus or minus 1. By multiplying by 1, equation (\ref{eq: odd even solution step 1 appendix}) can be manipulated into the form
\begin{equation}
\centering
\eqalign{\frac{E_s}{s}e^{-in\pi/(2r)}[(\tilde{x_\nu}e^{in\pi/r}e^{-in\pi(d-1)/(rs)})^s e^{in\pi(d-s) /r}\\
+{e^{in\pi/r}e^{-in\pi(d-s) /r}}{(\tilde{x_\nu}e^{in\pi/r}e^{-in\pi(d-1)/(rs)})^{-s}}] \\
+\frac{E_r}{r}[(\tilde{x_\nu}e^{in\pi/r}e^{-in\pi(d-1)/(rs)})^r e^{-in\pi/r}e^{in\pi(d-1)/(s)}\\
+{e^{in\pi/r}e^{-in\pi(d-1)/(s)}}{(\tilde{x_\nu}e^{in\pi}e^{-in\pi(d-1)/(rs)})^{-r}}] \\
- 2\omega\left[p_{0\parallel} \pm i\sqrt{2I_p+p_{0\perp}^2}\right] = 0.}\label{eq: odd even solution step 2 appendix} \
\end{equation}
where $d, k$ and $q$ are a solution to the Diophantine equations
\begin{eqnarray}
\centering
d - 1 &=& sq\label{eq: diophantine 1 appendix} \\
d - s &=& rk\label{eq: diophantine 2 appendix} \
\end{eqnarray}

By using the parities of $r$ and s to determine the parities of $k$ and $q$ for the cases $r$ odd, $s$ even and$ r$ even, $s$ odd separately, we can determine in each case whether $e^{i\pi n(d-1)/s}$ and $e^{i\pi n(d-1)/s}$ are plus or minus one. By adjusting the form of the solution, (\ref{eq: odd even solution step 2 appendix}) can be returned to the form of (\ref{eq: general two colour spe}). The transformation of the solution required to do this is given by equation (\ref{eq: odd even solution transformation appendix}).
\begin{eqnarray}
\centering
x_\mu(p_{0\parallel}) = (-1)^{n(p+1)}(-1)^{nq}\tilde{x_\nu}((-1)^n p_{0\parallel})e^{i\pi n/r}e^{-i\pi n(d-1)/(rs)}.\label{eq: odd even solution transformation appendix} \
\end{eqnarray}
This can be returned to an expression which includes the saddle point time, by making the substitution $x_\nu=\exp(i\omega t'_\nu)$. After some manipulations, we obtain 

\begin{eqnarray}
\centering
\mathrm{Re} [t'_\mu(p_{0\parallel})] &=& \frac{T}{2}\left[\frac{n}{r} - \frac{n(d-1)}{rs} + n(q+k+1)\right]-\mathrm{Re} [t'_\nu((-1)^n p_{0\parallel})] \label{eq: odd even real time solution transformation appendix} \\
\mathrm{Im} [t'_\mu(p_{0\parallel})] &=& \mathrm{Im} [t'_\nu((-1)^n p_{0\parallel})],\label{eq: odd even imag time solution transformation appendix} \
\end{eqnarray}
where $\mu$, $\nu$ are integer numbers. 
Equation (\ref{eq: odd even real time solution transformation appendix}) and (\ref{eq: odd even imag time solution transformation appendix}) can be simplified further by solving the diophantine equations (\ref{eq: diophantine 1 appendix}) and (\ref{eq: diophantine 2 appendix}) in terms of the multiplicative inverse of $r$ modulo $s$, $r_s^{-1}$. 
\begin{eqnarray}
\centering
\mathrm{Re} [t_\mu(p_{0\parallel})] &=& \frac{nT}{2}\left[r_s^{-1} + C\left(\frac{1}{r}-1\right)\right]-\mathrm{Re} [t_\nu((-1)^n p_{0\parallel})] \label{eq: odd even real time solution transformation simple appendix} \\
\mathrm{Im} [t_\mu(p_{0\parallel})] &=& \mathrm{Im} [t_\nu((-1)^n p_{0\parallel})].\label{eq: odd even imag time solution transformation simple appendix} \
\end{eqnarray}
where $C = ({1-r_s^{-1}r})/{s}$.

Similar arguments can be made for $(r, s)$ which are both odd.  However, in this case we additionally have the half cycle symmetry for all $\phi$ and the simple transformation
\begin{eqnarray}
\centering
x_\mu(p_{0\parallel}) = -x_\nu(-p_{0\parallel})\label{eq: odd half cycle solution transformation appendix} \
\end{eqnarray}
is also a solution to equation (\ref{eq: general two colour spe}).
In terms of saddle point times equation (\ref{eq: odd half cycle solution transformation appendix}) becomes
\begin{eqnarray}
\centering
\mathrm{Re} [t'_\mu(p_{0\parallel})] &=& \frac{T}{2} + \mathrm{Re} [t'_\nu(-p_{0\parallel})] \label{eq: odd half cycle real time solution transformation appendix} \\
\mathrm{Im} [t'_\mu(p_{0\parallel})] &=& \mathrm{Im} [t'_\nu(-p_{0\parallel})]\label{eq: odd half cycle imag time solution transformation appendix} \
\end{eqnarray}

Now, for $\phi = {n\pi}/{2s}$ such that $n \in 2\mathbb{Z}$ the following transformation gives another solution of equation (\ref{eq: general two colour spe}), namely 
\begin{eqnarray}
\centering
x_\mu(p_{0\parallel}) = (-1)^{n/2}\tilde{x_\nu}((-1)^{n/2} p_{0\parallel})e^{i\pi n/r}e^{-i\pi n(d-1)/(rs)}.\label{eq: odd solution transformation appendix} \
\end{eqnarray}
This is found in the same way as the $(r,s)$ opposite parity case by multiplying by one and introducing the same $d$, $k$,$\textrm{ and } q$ as before. This strategy can be used to define the transformation of $x_\nu$ in equation (\ref{eq: odd solution transformation appendix}), which leaves equation (\ref{eq: general two colour spe}) invariant.
This corresponds to the saddle point times having the symmetries 
\begin{eqnarray}
\centering
\mathrm{Re} [t'_\mu(p_{0\parallel})] &=& \frac{T}{2}\left[\frac{n}{r} - \frac{n(d-1)}{rs} + \frac{n}{2} \right]-\mathrm{Re} [t'_\nu((-1)^{n/2} p_{0\parallel})] \label{eq: odd real time solution transformation appendix} \\
\mathrm{Im} [t'_\mu(p_{0\parallel})] &=& \mathrm{Im} [t'_\nu((-1)^{n/2} p_{0\parallel})]\label{eq: odd imag time solution transformation appendix} \
\end{eqnarray}
which as before can be simplified to
\begin{eqnarray}
\centering
\mathrm{Re} [t'_\mu(p_{0\parallel})] &=& \frac{nT}{2}\left[\frac{C}{r} + \frac{1}{2} \right]-\mathrm{Re} [t'_\nu((-1)^{n/2} p_{0\parallel})] \quad \textrm{and} \label{eq: odd real time solution transformation simple appendix} \\
\mathrm{Im} [t'_\mu(p_{0\parallel})] &=& \mathrm{Im} [t'_\nu((-1)^{n/2} p_{0\parallel})].\label{eq: odd imag time solution transformation simple appendix} \
\end{eqnarray}
\printbibliography
\end{document}